\documentclass[12pt]{article}
\usepackage{myart}

\normalbaselines
\input tcilatex

\begin{document}

\title{Hydrodynamic equations for incompressible inviscid fluid in terms of
generalized stream function}
\author{Yuri A.Rylov}
\date{Institute for Problems in Mechanics, Russian Academy of Sciences,\\
101-1, Vernadskii Ave., Moscow, 119526, Russia.\\
e-mail: rylov@ipmnet.ru}
\maketitle

\begin{abstract}
Hydrodynamic equations for ideal incompressible fluid are written in terms
of generalized stream function. Two-dimensional version of these equations
is transformed to the form of one dynamic equation for the stream function.
This equation contains arbitrary function which is determined by inflow
conditions given on the boundary. To determine unique solution, velocity and
vorticity (but not only velocity itself) must be given on the boundary. This
unexpected circumstance may be interpreted in the sense that the fluid has
more degrees of freedom, than it was believed. Besides, the vorticity is
less observable quantity as compared with the velocity. It is shown that the
Clebsch potentials are used essentially at the description of vortical flow
\end{abstract}



\newpage

\section{Introduction}

In this paper we write hydrodynamic equations for ideal fluid in terms of
Clebsch potentials {\cite{C57,C59}}. This representation admits one to
integrate hydrodynamic equations and to obtain effective description of
rotational stream flow. We have discovered unexpected fact that the
effective description of the rotational flow cannot be carried out without
introduction of the Clebsch potentials. In general, the problem of
irrotational stream flow and that of rotational stream flow are stated
differently. Let us show this in example of stationary two-dimensional flow
of incompressible fluid. In this case one of Clebsch potentials may coincide
with the stream function $\psi $. The irrotational flow is described by the
equation for stream function $\psi $%
\begin{equation}
\psi _{xx}+\psi _{yy}=0  \label{h1.1}
\end{equation}
where indices mean corresponding partial derivatives. For the rotational
stationary two-dimensional flow the equation (\ref{h1.1}) transforms to the
form 
\begin{equation}
\psi _{xx}+\psi _{yy}=\Omega \left( \psi \right)  \label{h1.1a}
\end{equation}
where $\Omega \left( \psi \right) $ is some function of argument $\psi $,
describing the character and intensity of vorticity. The obtained equation
distinguishes from (\ref{h1.1}) in additional term. Both equations (\ref
{h1.1}) and (\ref{h1.1a}) are the elliptic type equations. To obtain a
unique solution of equation (\ref{h1.1}), it is sufficient to give value of
the stream function $\psi $ on the boundary $\Sigma $ of the volume $V$,
where the flow is considered. The same is valid for equation (\ref{h1.1a}),
provided the form of the function $\Omega \left( \psi \right) $ is known.
But the value of $\psi $ on $\Sigma $ does not admit one to determine the
form of the function $\Omega \left( \psi \right) $. For determination of $%
\Omega \left( \psi \right) $ one needs some additional information, given on
the boundary $\Sigma $.

To state the problem of the rotational stream flow, it is necessary to
determine what kind of information on the boundary is sufficient for
derivation of the unique solution of equation (\ref{h1.1a}). Theory of
rotational flows cannot exist without statement of this problem.
Unfortunately, we have not seen in literature a statement of the problem of
the rotational stream flow, we know only single exact solutions \cite{L72}.
It means that the theory of rotational flows does not exist.

Equation (\ref{h1.1}) is a special case of equation (\ref{h1.1a}), when $%
\Omega \left( \psi \right) =0$. Equation (\ref{h1.1a}) is obtained from the
conventional hydrodynamic equations for two-dimensional stationary flow of
incompressible fluid, which have the form 
\begin{equation}
u_{x}+v_{y}=0,\qquad \partial _{y}\left( uu_{x}+vu_{y}\right) =\partial
_{x}\left( uv_{x}+vv_{y}\right)  \label{h1.2a}
\end{equation}
where $u$ and $v$ are velocity components along $x$-axis and $y$-axis
respectively. Introducing the stream function $\psi $%
\begin{equation}
u=-\psi _{y},\qquad v=\psi _{x}  \label{h1.2b}
\end{equation}
we satisfy the first equation (\ref{h1.2a}) identically. The second equation
(\ref{h1.2a}) takes the form 
\begin{equation}
\frac{\partial \left( \omega ,\psi \right) }{\partial \left( x,y\right) }%
=0,\qquad \omega \equiv \psi _{xx}+\psi _{yy}  \label{h1.2c}
\end{equation}
Relation (\ref{h1.1a}) is the general solution $\omega =\Omega \left( \psi
\right) $ of (\ref{h1.2c}), where $\Omega $ is an arbitrary function of $%
\psi $.

The goal of the present paper is the statement of the problem of the
rotational stream flow. The problem is solved effectively in terms of the
generalized stream function (GSF) which has several components. GSF is a
generalization of the usual stream function $\psi $ for two-dimensional flow
onto a more complicated $n$-dimensional case. Unfortunately, in other more
complicated cases the statement of the problem of the rotational flow needs
a special well developed technique.

This technique is based on application of Jacobian technique to the
descriptions of hydrodynamic equations written in terms of Clebsch
potentials. We shall refer to this technique as the generalized stream
function technique (GSF-technique). Use of Jacobian technique in application
to Clebsch potentials goes back to papers by Clebsch \cite{C57,C59}, who
obtained his potentials dealing with Jacobians. In contemporary
hydrodynamics the Clebsch potentials are considered formally without
connection with the Jacobian technique \cite{L32}. Clebsch potentials have
also another names (Lagrangian coordinates, Lagrangian variables \cite{ZK97}%
, labels of fluid particles). There are several version of representation of
Clebsch potentials. Elements of Jacobian technique was used by many authors
( \cite{H55,E60,B83,S88,ZK97} and many others). But in solution of the
problem of streamline flow the Jacobian technique is not used practically,
as well as Clebsch potentials. We are interested only in that version, which
is connected with a use of Jacobian technique (GSF-technique). We use
space-time symmetric version of the Jacobian technique which appears to be
simple and effective. It seems that the progress in the investigation of
vortical flows is connected mainly with the developed Jacobian technique
used together with Clebsch potentials (GSF-technique). Presentation of the
GSF-technique can be found in \cite{R999}, as well as in \cite{R99}, where
it is used for description of the fluid flow in terms of the wave functions.

Why are Clebsch potentials important in the statement of the problem on
vortical stream flow? The term $\Omega \left( \psi \right) $ in rhs of (\ref
{h1.1a}) describes vorticity. According to the Kelvin's theorem on
circulation the vorticity is ''frozen in the fluid'', and vorticity travels
together with the fluid. Clebsch potentials (labels of the fluid particles),
as well as the generalized stream function (GSF) are also ''frozen in the
fluid''. They also travel together with the fluid and vorticity. As a result
the Clebsch potentials (labels) and the generalized stream function (GSF)
appear to be an effective tool of the vorticity description.

A rotational fluid flow has some properties which are absent in irrotational
flows.

\begin{enumerate}
\item  Consideration of the fluid displacement and a use of Clebsch
potentials describing this displacement is essential in rotational flows,
because this displacement transfers vorticity, which influences the velocity
field, whereas such a transport is of no importance in irrotational flows,
where the vorticity vanishes.

\item  The boundary conditions for a rotational flow contain more
information, than the boundary conditions for irrotational flow. This
additional information is information on vorticity, which conserves in any
ideal fluid. Although the field of vorticity 
\[
\mathbf{\omega }=\mathbf{\nabla \times v} 
\]
is determined by the velocity field $\mathbf{v}$, but it is valid only
inside some 3-volume $V$, where vorticity can be determined as a result of
differentiation of the velocity field. On the boundary $\Sigma $ of 3-volume 
$V$ one can calculate only component of $\mathbf{\omega }$ along the normal
to $\Sigma .$ Components of vorticity $\mathbf{\omega }$ tangent to $\Sigma $
must be given additionally. It means that additional (as compared with the
irrotational flow) degrees of freedom appear in the rotational flow, and
additional information in boundary conditions is necessary for their
description.
\end{enumerate}

The first property is analyzed in \cite{R999}. We present it here also. It
is common practice to think that the problem of streamline flow can be
solved by consideration of only Euler system of equations 
\begin{equation}
\frac{\partial \rho }{\partial t}+\nabla (\rho \mathbf{v})=0  \label{b1.1}
\end{equation}
\begin{equation}
\frac{\partial \mathbf{v}}{\partial t}+(\mathbf{v\nabla })\mathbf{v}=-\frac{1%
}{\rho }\mathbf{\nabla }p,\qquad p=\rho ^{2}\frac{\partial E}{\partial \rho }
\label{b1.2}
\end{equation}
\begin{equation}
\frac{\partial S}{\partial t}+(\mathbf{v\nabla })S=0  \label{b1.3}
\end{equation}
where $p$ is the pressure, and $E=E(\rho ,S)$ is the internal energy of an
unit mass considered to be a function of the mass density $\rho $ and the
entropy $S$. The internal energy $E=E(\rho ,S)$ is an unique characteristic
of the ideal fluid. Displacement of the fluid particles (i.e. their
trajectories and the law of motion along them) in the given velocity field $%
\mathbf{v}$ is described by the equations 
\begin{equation}
\frac{d\mathbf{x}}{dt}=\mathbf{v}(t,\mathbf{x}),\qquad \mathbf{x}=\mathbf{x}%
(t,\mathbf{\xi })  \label{b1.6}
\end{equation}
where $\mathbf{v}(t,\mathbf{x})$ is a solution of the system (\ref{b1.1})-(%
\ref{b1.3}). The equations (\ref{b1.1}) - (\ref{b1.6}) form the complete
system of hydrodynamic equations, but the form of this system is not quite
consistent, because dynamic variables $\rho ,\mathbf{v},S$ in (\ref{b1.1}) -
(\ref{b1.3}) are functions of $t,\mathbf{x}$, whereas $\mathbf{x}$ in (\ref
{b1.6}) is a function of $t$ and of the label $\mathbf{\xi }$ of the fluid
particle.

Let $\mathbf{\xi }=\mathbf{\xi }(t,\mathbf{x})=\{\xi _{\alpha }(t,\mathbf{x}%
)\}$, $\alpha =1,2,3$ be three independent integrals of equations (\ref{b1.6}%
). Values $\mathbf{\xi }$ of three integrals can label fluid particles
(Lagrangian coordinates). Then labels $\mathbf{\xi }$ satisfy three
equations 
\begin{equation}
\frac{\partial \mathbf{\xi }}{\partial t}+(\mathbf{v\nabla })\mathbf{\xi }=0,
\label{b1.4}
\end{equation}
which are equivalent to three equations (\ref{b1.6}). The system of
equations (\ref{b1.1}) - (\ref{b1.6}) is hardly perceived as a whole dynamic
system, because the system (\ref{b1.1})-(\ref{b1.3}) is closed and its
dynamic variables $\rho ,\mathbf{v},S$ are functions of $t,\mathbf{x}$,
whereas the dynamic variables of (\ref{b1.6}) are functions of variables $t,%
\mathbf{\xi }$\textbf{. }But equivalent system (\ref{b1.1})-(\ref{b1.3}), (%
\ref{b1.4}) of equations for variables $\rho ,\mathbf{v},S,\mathbf{\xi }$,
which depend on $t,\mathbf{x}$, is the whole system of dynamic equations.
This system is obtained from the variational principle, whereas the Euler
system (\ref{b1.1})-(\ref{b1.3}) of hydrodynamic equations can be obtained
from the variational principle only for the case of irrotational flow. The
system (\ref{b1.1})-(\ref{b1.3}) is a closed subsystem of the whole system
of dynamic equations (\ref{b1.1})-(\ref{b1.3}), (\ref{b1.4}). On the
foundation of closure the Euler system (\ref{b1.1})-(\ref{b1.3}) is
considered conventionally as the complete (full) system of hydrodynamic
equations, whereas in reality the Euler system (\ref{b1.1})-(\ref{b1.3}) is
only a curtailed system, i.e. only a part of the complete (full) system of
hydrodynamic equations formed by equations (\ref{b1.1})-(\ref{b1.3}), (\ref
{b1.4}). If we work with the Euler system only, we cannot integrate it, in
general. If nevertheless we integrate it in some special cases, the
arbitrary functions of Clebsch potentials $\mathbf{\xi }$ appear in
integrated dynamic equations. If we use the complete system (\ref{b1.1})-(%
\ref{b1.3}), (\ref{b1.4}), we can always integrate it and reduce the number
of dynamic equations. This integration is accompanied by appearance of three
arbitrary functions $\mathbf{g}\left( \mathbf{\xi }\right) =\left\{
g^{\alpha }\left( \mathbf{\xi }\right) \right\} ,\;\alpha =1,2,3$ of Clebsch
potentials $\mathbf{\xi }$ in dynamic equations. These functions contain
full information on initial and boundary conditions for the fluid flow.

The integrated dynamic equation have different form for the irrotational and
rotational flows. In particular, if $\mathbf{\nabla }\times \mathbf{g}=0$,
what corresponds to irrotational flow of barotropic fluid, the equations (%
\ref{b1.4}), known as Lin constraints \cite{L63}, are not dynamic equations
in the integrated system of dynamic equations, and the variables $\mathbf{%
\xi }$ have no relation to calculation of the fluid flow. In general case of
rotational flow, when $\mathbf{\nabla }\times \mathbf{g}\neq 0$ identically,
the equations (\ref{b1.4}) are dynamic equations for the fluid. We show
this, as soon as the corresponding mathematical technique will be developed.

In the second section Jacobian technique is considered. The third section is
devoted to derivation of hydrodynamic equations of ideal fluid from the
variational principle. The variational principle for incompressible fluid is
considered in the fourth section. In the fifth section two-dimensional flow
of incompressible fluid is described in terms of the generalized stream
function. In the sixth section the inflow conditions for the stationary
two-dimensional flow of incompressible fluid are introduced. The seventh
section is devoted to stationary flow around an obstacle. Some examples of
two-dimensional stationary flow of incompressible fluid are considered in
the eighth section. 

\section{Jacobian technique}

Let us consider such a space-time symmetric mathematical object as the
Jacobian 
\begin{equation}
J\equiv \frac{\partial (\xi _{0},\xi _{1},\xi _{2},\xi _{3})}{\partial
(x^{0},x^{1},x^{2},x^{3})}\equiv \det \left| \left| \xi _{i,k}\right|
\right| ,\qquad \xi _{i,k}\equiv \partial _{k}\xi _{i}\equiv \frac{\partial
\xi _{i}}{\partial x^{k}},\qquad i,k=0,1,2,3  \label{b2.1}
\end{equation}
Here $\xi =\{\xi _{0},\mathbf{\xi }\}=\{\xi _{0},\xi _{1},\xi _{2},\xi
_{3}\} $ are four scalar considered to be functions of $x=\{x^{0},\mathbf{x}%
\}$, $\xi =\xi (x)$. The functions $\{\xi _{0},\xi _{1},\xi _{2},\xi _{3}\}$
are supposed to be independent in the sense that $J\neq 0$. It is useful to
consider the Jacobian $J$ as 4-linear function of variables $\xi
_{i,k}\equiv \partial _{k}\xi _{i}$, $i,k=0,1,2,3$. Then one can introduce
derivatives of $J$ with respect to $\xi _{i,k}$. The derivative $\partial
J/\partial \xi _{i,k}$ appears as a result of a replacement of $\xi _{i}$ by 
$x^{k}$ in the relation (\ref{b2.1}).

\begin{equation}
\frac{\partial J}{\partial \xi _{i,k}}\equiv \frac{\partial (\xi _{0},...\xi
_{i-1},x^{k},{\xi _{i+1},...}\xi _{3})}{\partial (x^{0},x^{1},x^{2},x^{3})}%
,\qquad i,k=0,1,2,3  \label{b2.2}
\end{equation}
For instance 
\begin{equation}
\frac{\partial J}{\partial \xi _{0,i}}\equiv \frac{\partial (x^{i},\xi
_{1},\xi _{2},\xi _{3})}{\partial (x^{0},x^{1},x^{2},x^{3})},\qquad i=0,1,2,3
\label{b2.3}
\end{equation}
This rule is valid for higher derivatives of $J$ also. 
\[
\frac{\partial ^{2}J}{\partial \xi _{i,k}\partial \xi _{s,l}}\equiv \frac{%
\partial (\xi _{0},...\xi _{i-1},x^{k},\xi _{i+1},...\xi _{s-1},x^{l},\xi
_{s+1},...\xi _{3})}{\partial (x^{0},x^{1},x^{2},x^{3})} 
\]
\begin{equation}
\equiv \frac{\partial (x^{k},x^{l})}{\partial (\xi _{i},\xi _{s})}\frac{%
\partial (\xi _{0},\xi _{1},\xi _{2},\xi _{3})}{\partial
(x^{0},x^{1},x^{2},x^{3})}\equiv J(\frac{\partial x^{k}}{\partial \xi _{i}}%
\frac{\partial x^{l}}{\partial \xi _{s}}-\frac{\partial x^{k}}{\partial \xi
_{s}}\frac{\partial x^{l}}{\partial \xi _{i}}),\qquad i,k,l,s=0,1,2,3
\label{b2.4}
\end{equation}
It follows from (\ref{b2.1}), (\ref{b2.2}) that 
\begin{eqnarray}
\frac{\partial x^{k}}{\partial \xi _{i}} &\equiv &\frac{\partial (\xi
_{0},...\xi _{i-1},x^{k},{\xi _{i+1},...}\xi _{3})}{\partial {(}\xi _{0},\xi
_{1},\xi _{2},\xi _{3}{)}}  \nonumber \\
&\equiv &\frac{\partial (\xi _{0},...\xi _{i-1},x^{k},\xi _{i+1},{...}\xi
_{3})}{\partial (x^{0},x^{1},x^{2},x^{3})}\frac{{\partial (}%
x^{0},x^{1},x^{2},x^{3}{)}}{{\partial {(}}\xi _{0},\xi _{1},\xi _{2},\xi _{3}%
{)}}\equiv \frac{1}{J}\frac{\partial J}{\partial \xi _{i,k}},  \label{b2.5}
\\
i,k &=&0,1,2,3  \nonumber
\end{eqnarray}
and (\ref{b2.4}) may be written in the form 
\begin{equation}
\frac{\partial ^{2}J}{\partial \xi _{i,k}\partial \xi _{s,l}}\equiv \frac{1}{%
J}(\frac{\partial J}{\partial \xi _{i,k}}\frac{\partial J}{\partial \xi
_{s,l}}-\frac{\partial J}{\partial \xi _{i,l}}\frac{\partial J}{\partial \xi
_{s,k}}),\qquad i,k,l,s=0,1,2,3  \label{b2.6}
\end{equation}

The derivative $\partial J/\partial \xi _{i,k}$ is a cofactor to the element 
$\xi _{i,k}$ of the determinant (\ref{b2.1}). Then one has the following
identities 
\begin{equation}
\xi _{l,k}\frac{\partial J}{\partial \xi _{s,k}}\equiv \delta
_{l}^{s}J,\qquad \xi _{k,l}\frac{\partial J}{\partial \xi _{k,s}}\equiv
\delta _{l}^{s}J,\qquad l,s=0,1,2,3  \label{b2.7}
\end{equation}
\begin{equation}
\partial _{k}\frac{\partial J}{\partial \xi _{i,k}}\equiv \frac{\partial
^{2}J}{\partial \xi _{i,k}\partial \xi _{s,l}}\partial _{k}\partial _{l}\xi
_{s}\equiv 0,\qquad i=0,1,2,3.  \label{b2.8}
\end{equation}
Here and in what follows a summation over two repeated indices is produced
(0-3) for Latin indices. The identity (\ref{b2.8}) can be considered as a
corollary of the identity (\ref{b2.6}) and a symmetry of $\partial
_{k}\partial _{l}\xi _{s}$ with respect to permutation of indices $k$, $l$.
Convolution of (\ref{b2.6}) with $\partial _{k}$, or $\partial _{l}$
vanishes also. 
\begin{equation}
\partial _{k}\frac{\partial ^{2}J}{\partial \xi _{i,k}\partial \xi _{s,l}}%
\equiv \frac{\partial ^{3}J}{\partial \xi _{i,k}\partial \xi _{s,l}\partial
\xi _{m,n}}\partial _{k}\partial _{n}\xi _{m}\equiv 0,\qquad i,l,s=0,1,2,3.
\label{b2.8a}
\end{equation}

Relations (\ref{b2.1}) --(\ref{b2.6}) are written for four independent
variables $x$, but they are valid in an evident way for arbitrary number of $%
n+1$ variables $x=\{x^{0},x^{1},\ldots x^{n}\}$ and $\xi =\{\xi _{0},\mathbf{%
\xi }\}$, \quad $\mathbf{\xi }=\{\xi _{1},\xi _{2},\ldots \xi _{n}\}$.

Application of the Jacobian $J$ to hydrodynamics is founded on the property,
which can be formulated as the property of the generalized stream function
(GSF) $\mathbf{\xi =}\{\xi _{1},\xi _{2},...,\xi _{n}\}$ in the $(n+1)$%
-dimensional space of coordinates $x=\left\{ x^{0},x^{1},...,x^{n}\right\} $.

On the basis of the generalized stream function $\mathbf{\xi }$ one can
construct $n$-vector $j^{i}$ 
\begin{equation}
j^{i}=m\frac{\partial J}{\partial \xi _{0,i}},\qquad j^{i}=\left\{ \rho
,\rho \mathbf{v}\right\} ,\qquad m=\text{const}  \label{b2.9}
\end{equation}
in such a way that $j^{i}$ satisfies the continuity equation 
\begin{equation}
\partial _{i}j^{i}=0  \label{b2.10}
\end{equation}
identically for any choice of variables $\mathbf{\xi }$. Besides the
variables $\mathbf{\xi }$ are constant along any line $\mathcal{L}$ tangent
to $n$-vector $j^{i}$, and can label this line, because the set of
quantities $\mathbf{\xi }$ is different for different lines $\mathcal{L}$.

In the case of dynamical system (\ref{b1.1}) - (\ref{b1.3}), (\ref{b1.4}) we
have $j^{i}=\left\{ \rho ,\rho \mathbf{v}\right\} $. It means that equations
(\ref{b1.1}), (\ref{b1.4}) are satisfied at any choice of the generalized
stream function (Clebsch potentials) $\mathbf{\xi }$. Substituting the flux
vector $j^{i}=\left\{ \rho ,\rho \mathbf{v}\right\} $, expressed via $%
\mathbf{\xi }$\textbf{,} in remaining hydrodynamic equations, we obtain
dynamic equations for determination of generalized stream function $\mathbf{%
\xi }$. This procedure is insignificant in the case of irrotational flow,
when Lin constraints (\ref{b1.4}) are of no importance. But it is a very
effective procedure in the case of rotational flow, because it reduces the
number of dynamic equations to be solved.

The continuity equation (\ref{b2.10}) is used without approximation in all
hydrodynamic models, and the change of variables $\{\rho ,\rho \mathbf{%
v\}\leftrightarrow \xi }$ described by (\ref{b2.9}) appears sometimes to be
useful. In particular, in the case of two-dimensional established flow of
incompressible fluid the generalized stream function $\mathbf{\xi }$ reduces
to one variable $\xi _{2}=\psi $, known as the stream function. In this case
there are only two essential independent variables $x^{0}=x$, $x^{1}=y$, and
the relations (\ref{b2.9}), (\ref{b2.10}) reduce to relations 
\begin{equation}
\rho ^{-1}j_{x}=u=-\frac{\partial \psi }{\partial y},\qquad \rho
^{-1}j_{y}=v=\frac{\partial \psi }{\partial x},\qquad \frac{\partial u}{%
\partial x}+\frac{\partial v}{\partial y}=0  \label{b2.11}
\end{equation}
Defining the stream line as a line tangent to the flux $j$ 
\begin{equation}
\frac{dx}{j_{x}}=\frac{dy}{j_{y}},  \label{b2.12}
\end{equation}
we deduce that the stream function is constant along the stream line,
because according to two first equations (\ref{b2.11}), $\psi =\psi (x,y)$
is an integral of the equation (\ref{b2.12}).

In the general case, when the space dimension is $n$ and $%
x=\{x^{0},x^{1},\ldots x^{n}\}$, $\xi =\{\xi _{0},\mathbf{\xi }\}$, $\mathbf{%
\xi }=\{\xi _{1},\xi _{2},\ldots \xi _{n}\}$, the quantities $\mathbf{\xi }%
=\{\xi _{\alpha }\}$, $\alpha =1,2,\ldots n$ are constant along the line $%
\mathcal{L}:\;x=x(\tau )$ tangent to the flux vector $j=\{j^{i}\}$, $%
i=0,1,\ldots n$ 
\begin{equation}
\mathcal{L}:\qquad \frac{dx^{i}}{d\tau }=j^{i}(x),\qquad i=0,1,\ldots n
\label{b2.14}
\end{equation}
where $\tau $ is a parameter along the line $\mathcal{L}$. This statement is
formulated mathematically in the form 
\begin{equation}
\frac{d\xi _{\alpha }}{d\tau }=j^{i}\partial _{i}\xi _{\alpha }=m\frac{%
\partial J}{\partial \xi _{0,i}}\partial _{i}\xi _{\alpha }=0,\qquad \alpha
=1,2,\ldots n  \label{b2.15}
\end{equation}
The last equality follows from the first identity (\ref{b2.7}) taken for $%
s=0 $, $l=1,2,\ldots n$

Interpretation of the line (\ref{b2.14}) tangent to the flux is different
for different cases. If $x=\{x^{0},x^{1},\ldots x^{n}\}$ contains only
spatial coordinates, the line (\ref{b2.14}) is a line in the usual space. It
is regarded as a stream line, and $\mathbf{\xi }$ can be interpreted as
quantities which are constant along the stream line (i.e. as a generalized
stream function). If $x^{0}$ is the time coordinate, the equation (\ref
{b2.14}) describes a line in the space-time. This line (known as a world
line of a fluid particle) determines a motion of the fluid particle.
Variables $\mathbf{\xi }=\{\xi _{1},\xi _{2},\ldots \xi _{n}\}$ which are
constant along the world line are different, generally, for different
particles. If $\xi _{\alpha }$, $\alpha =1,2,\ldots n$ are independent, they
may be used for the fluid particle labeling. When one of coordinates $x$ is
timelike the set of variables $\mathbf{\xi }$ is not perceived as a
generalization of the stream function $\psi $. Nevertheless, we shall use
the term ''generalized stream function'' in all cases, because from
mathematical viewpoint it is of no importance whether coordinate $x^{0}$ is
timelike or spacelike.

Thus, although interpretation of the relation (\ref{b2.9}) considered as a
change of dependent variables $j$ by $\mathbf{\xi }$ may be different, from
the mathematical viewpoint this transformation means a replacement of the
continuity equation by some equations for the generalized stream function $%
\mathbf{\xi }$. Difference of the interpretation is of no importance in this
context.

Note that the expressions 
\begin{equation}
j^{i}=m\rho _{0}(\mathbf{\xi })\frac{\partial J}{\partial \xi _{0,i}}\equiv
m\rho _{0}(\mathbf{\xi })\frac{\partial (x^{i},\xi _{1},\xi _{2},\xi _{3})}{%
\partial (x^{0},x^{1},x^{2},x^{3})},\qquad i=0,1,2,3,  \label{b2.16}
\end{equation}
can be also considered as four-flux satisfying the continuity equation (\ref
{b2.10}). Here $m$ is a constant and $\rho _{0}(\mathbf{\xi })$ is an
arbitrary function of $\mathbf{\xi }$. It follows from the identity 
\begin{equation}
m\rho _{0}(\mathbf{\xi })\frac{\partial (x^{i},\xi _{1},\xi _{2},\xi _{3})}{%
\partial (x^{0},x^{1},x^{2},x^{3})}\equiv m\frac{\partial (x^{i},\tilde{\xi}%
_{1},\xi _{2},\xi _{3})}{\partial (x^{0},x^{1},x^{2},x^{3})},\qquad \tilde{%
\xi}_{1}=\int\limits_{0}^{\xi _{1}}\rho _{0}(\xi _{1}^{\prime },\xi _{2},\xi
_{3})d\xi _{1}^{\prime }.  \label{b2.17}
\end{equation}

The Jacobian technique is very useful for manipulation with hydrodynamic
equations (\ref{b1.1})-(\ref{b1.3}), (\ref{b1.4}). For instance, one can
integrate the complete system of 5+3 dynamic equations (\ref{b1.1})-(\ref
{b1.3}), (\ref{b1.4}) reducing it to the system of five dynamic equations
written in the form \cite{R999}

\begin{equation}
S(t,\mathbf{x})=S_{0}(\mathbf{\xi })  \label{b1.9}
\end{equation}
\begin{equation}
\rho (t,\mathbf{x})=\rho _{0}(\mathbf{\xi })\frac{\partial (\xi _{1},\xi
_{2},\xi _{3})}{\partial (x^{1},x^{2},x^{3})}\equiv \rho _{0}(\mathbf{\xi }){%
\frac{\partial (\mathbf{\xi })}{\partial (\mathbf{x})}}  \label{b1.10}
\end{equation}
\begin{equation}
\mathbf{v}(t,\mathbf{x})=\mathbf{u}(\varphi ,\mathbf{\xi },\eta ,S)\equiv
\nabla \varphi +g^{\alpha }(\mathbf{\xi })\nabla \xi _{\alpha }-\eta \nabla
S,  \label{b1.11}
\end{equation}
where $S_{0}(\mathbf{\xi })$, $\rho _{0}(\mathbf{\xi })$ and $g(\mathbf{\xi }%
)=\{g^{\alpha }(\mathbf{\xi })\}$, $\alpha =1,2,3$ are arbitrary integration
functions of argument $\mathbf{\xi }$. The quantities $\varphi $, $\eta $
are new dependent variables satisfying dynamic equations 
\begin{equation}
\frac{\partial \varphi }{\partial t}+\mathbf{u}(\varphi ,\mathbf{\xi },\eta
,S)\nabla \varphi -{\frac{1}{2}}{[\mathbf{u}(}\varphi {,\mathbf{\xi },}\eta
,S{)]}^{2}+\frac{\partial (\rho E)}{\partial \rho }=0  \label{b1.12}
\end{equation}
\begin{equation}
\frac{\partial \eta }{\partial t}+\mathbf{u}(\varphi ,\mathbf{\xi },\eta ,S)%
\mathbf{\nabla }\eta =-\frac{\partial E}{\partial S}\text{.}  \label{b1.13}
\end{equation}
If five dependent variables $\varphi $, $\mathbf{\xi }$, $\eta $ satisfy the
system of equations (\ref{b1.4}), (\ref{b1.12}), (\ref{b1.13}), five dynamic
variables $S$, $\rho $, $\mathbf{v}$ (\ref{b1.9})--(\ref{b1.11}) satisfy
dynamic equations (\ref{b1.1})--(\ref{b1.3}). Indefinite functions $S_{0}(%
\mathbf{\xi })$, $\rho _{0}(\mathbf{\xi })$, and $g^{\alpha }(\mathbf{\xi })$
can be determined from initial and boundary conditions in such a way that
the initial and boundary conditions for variables $\varphi $, $\mathbf{\xi }$%
, $\eta $ were universal in the sense that they do not depend on the fluid
flow \cite{R999}. Further dynamic equations (\ref{b1.4}), (\ref{b1.12}), (%
\ref{b1.13}) will be derived directly from the variational principle.

\section{ Variational principle}

The ideal (nondissipative) fluid is a continuous dynamical system whose
dynamic equations can be derived from the variational principle with the
action functional 
\begin{equation}
\mathcal{A}_{\mathrm{L}}[\mathbf{x}]=\int \{\frac{m}{2}(\frac{d\mathbf{x}}{dt%
})^{2}-V\}\rho _{0}(\mathbf{\xi })dtd\mathbf{\xi },  \label{b3.1}
\end{equation}
where the fluid particle coordinates $\mathbf{x}=\{x^{\alpha }(t,\mathbf{\xi 
})\}$, $\alpha =1,2,3$ are dependent variables considered to be functions of
time $t$ and of labels (Lagrangian coordinates) $\mathbf{\xi }=\{\xi
_{1},\xi _{2},\xi _{3}\}$. $d\mathbf{x}/dt$ is a derivative of $\mathbf{x}$
with respect to $t$ taken at fixed $\mathbf{\xi }$. 
\begin{equation}
\frac{dx^{\alpha }}{dt}\equiv \frac{\partial (x^{\alpha },\xi _{1},\xi
_{2},\xi _{3})}{\partial (t,\xi _{1},\xi _{2},\xi _{3})}\equiv \frac{%
\partial (x^{\alpha },\mathbf{\xi })}{\partial (t,\mathbf{\xi })}\qquad
\alpha =1,2,3  \label{b3.2}
\end{equation}
$\rho _{0}(\mathbf{\xi })$ is some non-negative weight function, and $V$ is
a potential of a self-consistent force field which depends on $\mathbf{\xi }$%
, $\mathbf{x}$ and derivatives of $\mathbf{x}$ with respect to $\mathbf{\xi }
$. $m=$const is a mass of the fluid particle. The value of mass is
unessential, and without loss of generality one may set $m=1$. For the ideal
fluid the potential $V$ is such a function of $\mathbf{\xi }$ and $\partial 
\mathbf{x/}\partial \mathbf{\xi }$ which can be represented in the form 
\begin{equation}
V(\mathbf{\xi },\partial \mathbf{x/}\partial \mathbf{\xi })=mE(\rho
,S),\qquad \rho =\rho _{0}(\mathbf{\xi })\left( \frac{\partial
(x^{1},x^{2},x^{3})}{\partial (\xi _{1},\xi _{2},\xi _{3})}\right)
^{-1},\qquad S=S_{0}(\mathbf{\xi })  \label{c3.2}
\end{equation}
where the entropy per unit mass $S=S_{0}(\mathbf{\xi })$ is a given function
of $\mathbf{\xi }$ determined by the initial conditions. $\rho $ is the
fluid density, and $E(\rho ,S)$ is the internal energy of the fluid per unit
mass of the fluid. The quantity $E(\rho ,S)$ is considered to be a given
function of its arguments.

The variational principle (\ref{b3.1}) generates the sixth order system of
dynamic equations for six dependent variables $\mathbf{x}$, $d\mathbf{x}/dt$
considered to be functions of independent variables $t$, $\mathbf{\xi }$.
Such a way of description is known as the Lagrangian description of the
fluid. If the variables $t$, $\mathbf{x}$ are considered to be independent
variables, and six variables $\mathbf{\xi }$, $\partial \mathbf{\xi /}%
\partial t$ are considered to be dependent variables, the sixth order system
of dynamic equations for $\mathbf{\xi }$, $\partial \mathbf{\xi /}\partial t$
arises. It is known as the Eulerian description of the fluid.

The partial choice of labeling of the fluid particles by variables $\mathbf{%
\xi }$ is not material from a physical point of view. This circumstance is
displayed in an existence of the relabeling transformation group 
\begin{equation}
\xi _{\alpha }\rightarrow \tilde{\xi}_{\alpha }=\tilde{\xi}_{\alpha }(%
\mathbf{\xi }),\qquad D\equiv \det \parallel \partial \tilde{\xi}_{\alpha
}/\partial \xi _{\beta }\parallel \neq 0,\qquad \alpha ,\beta =1,2,3
\label{b1.16}
\end{equation}
The action (\ref{b3.1}) appears to be invariant with respect to the
relabeling group, provided $V$ has the form (\ref{c3.2}), where $S$ is a
scalar and $\rho (\mathbf{\xi })$ is a scalar density, i.e. under the
transformations (\ref{b1.16}) $\rho _{0}(\mathbf{\xi })$ transforms as
follows 
\begin{equation}
\rho _{0}(\mathbf{\xi })\rightarrow \tilde{\rho}_{0}(\tilde{\mathbf{\xi }}%
),\qquad \rho (\mathbf{\xi })\rightarrow \tilde{\rho}(\tilde{\mathbf{\xi }}%
)=\rho (\mathbf{\xi })D  \label{b3.6}
\end{equation}
The relabeling group is used in hydrodynamics comparatively recently \cite
{E38,E60,C63,B70,FS78,S82,ZK97}.

The relabeling group is a symmetry group of the dynamical system. It may be
used to simplify a description of the fluid. There are at least two
different ways of usage of the relabeling group. The first (conventional)
way of such a simplification is a use of five relabeling-invariant variables 
$\mathbf{v}=d\mathbf{x}/dt,$ $\rho ,$ $S$ considered to be functions of
independent variables $t,\mathbf{x}$. This way of description will be
referred to as the relabeling invariant description (RID). The variables $%
\mathbf{\xi }$, describing labeling, are eliminated. At this elimination of $%
\mathbf{\xi }$ the dynamic equation (\ref{b1.4}) is eliminated also on the
foundation that the remaining Eulerian system of dynamic equations (\ref
{b1.1})-(\ref{b1.3}) for five dependent dynamic variables $\mathbf{v},$ $%
\rho ,$ $S$ is closed. Conventionally most researchers use RID, and hence
the Eulerian system of dynamic equations. Another way of simplification is a
use of the relabeling group for integration of the complete system of
dynamic equations (\ref{b1.1})-(\ref{b1.3}), (\ref{b1.4}) and application of
the system (\ref{b1.4}), (\ref{b1.12}), (\ref{b1.13}).

In general, equivalency of the system (\ref{b1.4}), (\ref{b1.12}), (\ref
{b1.13}) and the system (\ref{b1.1})-(\ref{b1.3}), (\ref{b1.4}) can be
verified by a direct substitution of variables $\rho $, $S$, $\mathbf{v}$,
defined by the relations (\ref{b1.9})--(\ref{b1.11}), into the equations (%
\ref{b1.1})--(\ref{b1.3}). Using equations (\ref{b1.4}), (\ref{b1.12}), (\ref
{b1.13}), one obtains identities after subsequent calculations. But such
computations do not display a connection between the integration and the
invariance with respect to the relabeling group (\ref{b1.16}). Besides, a
meaning of new variables $\varphi $, $\eta $ is not clear.

We shall use for our investigations the variational principle (\ref{b3.1}).
Note that for a long time a derivation of a variational principle for
hydrodynamic equations (\ref{b1.1})--(\ref{b1.3}) existed as an independent
problem \cite{D49,H55,E60,L63,SW67,B70,S88}. Existence of this problem was
connected with a lack of understanding that the system of hydrodynamic
equations (\ref{b1.1})--(\ref{b1.3}) is a curtailed system, and the full
system of dynamic equations (\ref{b1.1})-(\ref{b1.3}), (\ref{b1.4}) includes
equations (\ref{b1.4}) describing a motion of the fluid particles in the
given velocity field. The variational principle can generate only the
complete system of dynamic variables (but not its closed subsystem). Without
understanding this one tried to form the Lagrangian for the system (\ref
{b1.1})--(\ref{b1.3}) as a sum of some quantities taken with Lagrange
multipliers. lhs of dynamic equations (\ref{b1.1})--(\ref{b1.3}) and some
other constraints were taken as such quantities.

Now this problem has been solved (see review by Salmon\cite{S88}) on the
basis of the Eulerian version of the variational principle (\ref{b3.1}),
where equations (\ref{b1.4}) appear automatically and cannot be ignored. In
our version of the variational principle we follow Salmon \cite{S88} with
some modifications which underline a curtailed character of hydrodynamic
equations (\ref{b1.1})--(\ref{b1.3}), because the understanding of the
curtailed character of the system (\ref{b1.1})--(\ref{b1.3}) removes the
problem of derivation of the variational principle for the hydrodynamic
equations (\ref{b1.1})--(\ref{b1.3}).

The starting point is the action (\ref{b3.1}). We prefer to work with
Eulerian description, when Lagrangian coordinates (particle labeling) $\xi
=\{\xi _{0},\mathbf{\xi }\}$, $\mathbf{\xi }=\{\xi _{1},\xi _{2},\xi _{3}\}$
are considered to be dependent variables, and Eulerian coordinates $%
x=\{x^{0},\mathbf{x}\}=\{t,\mathbf{x}\}$, $\mathbf{x}=\{x^{1},x^{2},x^{3}\}$
are considered to be independent variables. Here $\xi _{0}$ is a temporal
Lagrangian coordinate which evolves along the particle trajectory in an
arbitrary way. Now $\xi _{0}$ is a fictitious variable, but after
integration of equations the variable $\xi _{0}$ ceases to be fictitious and
turns into the variable $\varphi $, appearing in the integrated system (\ref
{b1.4}), (\ref{b1.12}), (\ref{b1.13}).

Further mainly space-time symmetric designations will be used, that
simplifies considerably all computations. In the Eulerian description the
action functional (\ref{b3.1}) is to be represented as an integral over
independent variables $x=\{x^{0},\mathbf{x}\}=\{t,\mathbf{x}\}$. We use the
Jacobian technique for such a transformation of the action (\ref{b3.1}),

Let us note that according to (\ref{b2.5}) the derivative (\ref{b3.2}) can
be written in the form 
\begin{equation}
v^{\alpha }=\frac{dx^{\alpha }}{dt}\equiv \frac{\partial J}{\partial \xi
_{0,\alpha }}\left( \frac{\partial J}{\partial \xi _{0,0}}\right)
^{-1},\qquad \alpha =1,2,3.  \label{b3.3}
\end{equation}
Then components of the 4-flux $j=\{j^{0},\mathbf{j}\}\equiv \{\rho ,\rho 
\mathbf{v}\}$ can be written in the form (\ref{b2.16}), provided the
designation (\ref{b1.10}) 
\begin{equation}
j^{0}=\rho =m\rho _{0}(\mathbf{\xi })\frac{\partial J}{\partial \xi _{0,0}}%
\equiv m\rho _{0}(\mathbf{\xi })\frac{\partial (x^{0},\xi _{1},\xi _{2},\xi
_{3})}{\partial (x^{0},x^{1},x^{2},x^{3})}  \label{b3.4}
\end{equation}
is used.

At such a form of the mass density $\rho $ the four-flux $j=\{j^{i}\}$, $%
i=0,1,2,3$ satisfies identically the continuity equation (\ref{b2.10}) which
takes place in virtue of identities (\ref{b2.7}), (\ref{b2.8}). Besides, in
virtue of identities (\ref{b2.7}), (\ref{b2.8}) the Lin constraints (\ref
{b1.4}) are fulfilled identically 
\begin{equation}
j^{i}\partial _{i}\xi _{\alpha }=m\rho _{0}(\mathbf{\xi })\frac{\partial J}{%
\partial \xi _{0,i}}\partial _{i}\xi _{\alpha }\equiv 0,\qquad \alpha =1,2,3.
\label{b3.5}
\end{equation}
Components $j^{i}$ are invariant with respect to the relabeling group (\ref
{b1.16}), provided the function $\rho _{0}(\mathbf{\xi })$ transforms
according to (\ref{b3.6}).

One has 
\begin{equation}
\rho _{0}(\mathbf{\xi })dtd\mathbf{\xi }=\rho _{0}(\mathbf{\xi })\frac{%
\partial J}{\partial \xi _{0,0}}dtd\mathbf{x}=\frac{\rho }{m}dtd\mathbf{x}
\label{b3.7}
\end{equation}
\begin{equation}
\frac{m}{2}\left( \frac{dx^{\alpha }}{dt}\right) ^{2}=\frac{m}{2}\left( 
\frac{\partial J}{\partial \xi _{0,\alpha }}\right) ^{2}\left( \frac{%
\partial J}{\partial \xi _{0,0}}\right) ^{-2},  \label{b3.8}
\end{equation}
and the variational problem with the action functional (\ref{b3.1}) is
written as a variational problem with the action functional 
\begin{equation}
\mathcal{A}_{\mathrm{E}}[\mathbf{\xi }]=\int \left\{ \frac{1}{2}\left( \frac{%
\partial J}{\partial \xi _{0,\alpha }}\right) ^{2}\left( \frac{\partial J}{%
\partial \xi _{0,0}}\right) ^{-2}-E\right\} \rho dtd\mathbf{x},  \label{b3.9}
\end{equation}
where $\rho $ is a fixed function of $\xi =\{\xi _{0},\mathbf{\xi }\}$ and
of $\xi _{\alpha ,i}\equiv \partial _{i}\xi _{\alpha }$, $\alpha =1,2,3$, $%
i=0,1,2,3$, defined by (\ref{b3.4}). $E$ is the internal energy of the fluid
which is supposed to be a fixed function of $\rho $ and $S_{0}(\mathbf{\xi }%
) $ 
\begin{equation}
E=E(\rho ,S_{0}(\mathbf{\xi })),  \label{b3.10}
\end{equation}
where $\rho $ is defined by (\ref{b3.4}) and $S_{0}(\mathbf{\xi })$ is some
fixed function of $\mathbf{\xi }$, describing initial distribution of the
entropy over the fluid.

The action (\ref{b3.9}) is invariant with respect to subgroup $\mathcal{G}%
_{S_0} $ of the relabeling group (\ref{b1.16}). The subgroup $\mathcal{G}%
_{S_0}$ is determined in such a way that any surface $S_0(\mathbf{\xi })=$%
const is invariant with respect to $\mathcal{G}_{S_0}$. In general, the
subgroup $\mathcal{G}_{S_0}$ is determined by two arbitrary functions of $%
\mathbf{\xi }$.

The action (\ref{b3.9}) generates the six order system of dynamic equations,
consisting of three second order equations for three dependent variables $%
\mathbf{\xi }$. Invariance of the action (\ref{b3.9}) with respect to the
subgroup $\mathcal{G}_{S_{0}}$ allow one to integrate the system of dynamic
equations. The order of the system becomes reduced, and two arbitrary
integration functions appear. The order of the system reduces to five (but
not to four), because the fictitious dependent variable $\xi _{0}$ ceases to
be fictitious as a result of the integration.

Unfortunately, the subgroup $\mathcal{G}_{S_0}$ depends on the form of the
function $S_0(\mathbf{\xi })$ and cannot be obtained in a general form. In
the special case, when $S_0(\mathbf{\xi })$ does not depend on $\mathbf{\xi }
$, the subgroup $\mathcal{G}_{S_0}$ coincides with the whole relabeling
group $\mathcal{G} $, and the order of the integrated system reduces to four.

In the general case it is convenient to introduce a new dependent variable 
\begin{equation}
S=S_{0}(\mathbf{\xi }).  \label{b3.11}
\end{equation}
Addition of the new variable increases the number of dynamic variables, but
at the same time this addition makes the action to be invariant with respect
to the whole relabeling group $\mathcal{G}$. It allows one to integrate the
dynamic system and to reduce the number of dynamic variables. According to (%
\ref{b3.5}) the variable $S$ satisfies the dynamic equation (\ref{b1.3}) 
\begin{equation}
j^{i}\partial _{i}S=0.  \label{b3.12}
\end{equation}
In virtue of designations (\ref{b2.16}) and identities (\ref{b2.7}), (\ref
{b2.8}) the equations (\ref{b3.12}), (\ref{b3.5}) are fulfilled identically.
Hence, they can be added to the action functional (\ref{b3.9}) as side
constraints without a change of the variational problem. Adding (\ref{b3.12}%
) to the Lagrangian of the action (\ref{b3.9}) by means of a Lagrange
multiplier $\eta $, one obtains 
\begin{equation}
\mathcal{A}_{\mathrm{E}}[\mathbf{\xi },\eta ,S]=\int \left\{ \frac{\rho }{2}%
\left( \frac{\partial J}{\partial \xi _{0,\alpha }}\right) ^{2}\left( \frac{%
\partial J}{\partial \xi _{0,0}}\right) ^{-1}-\rho E+\eta \frac{\partial J}{%
\partial \xi _{0,k}}\partial _{k}S\right\} dtd\mathbf{x}  \label{b3.13}
\end{equation}
where $E=E(\rho ,S)$. The action (\ref{b3.13}) is invariant with respect to
the relabeling group $\mathcal{G}$ which is determined by three arbitrary
functions of $\mathbf{\xi }$.

To obtain the dynamic equations, it is convenient to introduce new dependent
variables $j^{i}$, defined by (\ref{b2.16}). Let us introduce the new
variables $j^{i}$ by means of designations (\ref{b2.16}) taken with the
Lagrange multipliers $p_{i}$, $i=0,1,2,3$. Consideration of (\ref{b2.16}) as
side constraints does not change the variational problem, because conditions
(\ref{b2.16}) are always compatible with dynamic equations, generated by the
action (\ref{b3.13}). Then the action (\ref{b3.13}) takes the form 
\begin{equation}
\mathcal{A}_{\mathrm{E}}[\rho ,\mathbf{j},\mathbf{\xi },p,\eta ,S]=\int
\left\{ {\frac{\mathbf{j}^{2}}{2\rho }}-\rho E-p_{k}\left( j^{k}-m\rho _{0}(%
\mathbf{\xi })\frac{\partial J}{\partial \xi _{0,k}}\right) +\eta
j^{k}\partial _{k}S\right\} dtd\mathbf{x}  \label{b3.14}
\end{equation}

For obtaining dynamic equations, the variables $\rho ,\mathbf{j},\mathbf{\xi 
},p,\eta ,S$ are to be varied. Let us eliminate the variables $p_{i}$ from
the action (\ref{b3.14}). Dynamic equations arising as a result of a
variation with respect to $\xi _{\alpha }$ have the form 
\begin{equation}
{\frac{\delta \mathcal{A}_{\mathrm{E}}}{\delta \xi _{\alpha }}}\equiv \hat{%
\mathcal{L}}_{\alpha }p=-m\partial _{k}\left( \rho _{0}(\mathbf{\xi })\frac{%
\partial ^{2}J}{\partial \xi _{0,i}\partial \xi _{\alpha ,k}}p_{i}\right) +m%
\frac{\partial \rho _{0}(\mathbf{\xi })}{\partial \xi _{\alpha }}\frac{%
\partial J}{\partial \xi _{0,k}}p_{k}=0,\qquad \alpha =1,2,3  \label{b3.15}
\end{equation}
where $\hat{\mathcal{L}}_{\alpha }$ are linear operators acting on variables 
$p=\{p_{i}\}$, $i=0,1,2,3$. These equations can be integrated in the form 
\begin{equation}
p_{i}=g^{0}(\xi _{0})\partial _{i}\xi _{0}+g^{\alpha }(\mathbf{\xi }%
)\partial _{i}\xi _{\alpha },\qquad i=0,1,2,3,  \label{b3.16}
\end{equation}
where $\xi _{0}$ is some new variable (temporal Lagrangian coordinate), $%
g^{\alpha }(\mathbf{\xi })$, $\alpha =1,2,3$ are arbitrary functions of the
label $\mathbf{\xi }$, $g^{0}(\xi _{0})$ is an arbitrary function of $\xi
_{0}$. The relations (\ref{b3.16}) satisfy equations (\ref{b3.15})
identically. Indeed, substituting (\ref{b3.16}) into (\ref{b3.15}) and using
identities (\ref{b2.6}), (\ref{b2.7}), we obtain 
\begin{equation}
-m\partial _{k}\left\{ \rho _{0}(\mathbf{\xi })\left[ \frac{\partial J}{%
\partial \xi _{\alpha ,k}}g^{0}(\xi _{0})-\frac{\partial J}{\partial \xi
_{0,k}}g^{\alpha }(\mathbf{\xi })\right] \right\} +m\frac{\partial \rho _{0}(%
\mathbf{\xi })}{\partial \xi _{\alpha }}Jg^{0}(\xi _{0})=0,\qquad \alpha
=1,2,3,  \label{b3.17}
\end{equation}
Differentiating braces and using identities (\ref{b2.8}), (\ref{b2.7}), one
concludes that (\ref{b3.17}) is an identity.

Setting for simplicity 
\begin{equation}
\partial _{k}\varphi =g^{0}(\xi _{0})\partial _{k}\xi _{0},\qquad k=0,1,2,3
\label{b3.18}
\end{equation}
we obtain 
\begin{equation}
p_{k}=\partial _{k}\varphi +g^{\alpha }(\mathbf{\xi })\partial _{k}\xi
_{\alpha },\qquad k=0,1,2,3  \label{b3.19}
\end{equation}

Note that integration of equations (\ref{b3.15}) by means of the Jacobian
technique and appearance of arbitrary functions $g^{\alpha }\left( \mathbf{%
\xi }\right) $ is a result of invariance of the action with respect to the
relabeling group (\ref{b1.16}).

Substituting (\ref{b3.19}) in (\ref{b3.14}), one can eliminate variables $%
p_{i}$, $i=0,1,2,3$ from the functional (\ref{b3.14}). The term $g^{\alpha }(%
\mathbf{\xi })\partial _{k}\xi _{\alpha }\partial J/\partial \xi _{0,k}$
vanishes, the term $\partial _{k}\varphi \partial J/\partial \xi _{0,k}$
makes no contribution to the dynamic equations. The action functional takes
the form 
\begin{equation}
\mathcal{A}_{\mathbf{g}}[\rho ,\mathbf{j},\mathbf{\xi },\eta ,S]=\int
\left\{ \frac{\mathbf{j}^{2}}{2\rho }-\rho E-j^{k}\left( \partial
_{k}\varphi +g^{\alpha }(\mathbf{\xi })\partial _{k}\xi _{\alpha }-\eta
\partial _{k}S\right) \right\} dtd\mathbf{x}  \label{b3.20}
\end{equation}
where $g^{\alpha }(\mathbf{\xi })$ are considered to be fixed functions of $%
\mathbf{\xi }$ which are determined from initial conditions. Varying the
action (\ref{b3.20}) with respect to $\varphi $, $\mathbf{\xi }$, $\eta $, $%
S $, $\mathbf{j}$, $\rho $, we obtain dynamic equations 
\begin{equation}
\delta \varphi :\qquad \partial _{k}j^{k}=0,  \label{b3.21}
\end{equation}
\begin{equation}
\delta \xi _{\alpha }:\qquad \Omega ^{\alpha \beta }j^{k}\partial _{k}\xi
_{\beta }=0,\qquad \alpha =1,2,3,  \label{b3.22}
\end{equation}
where $\Omega ^{\alpha \beta }$ is defined by the relations 
\begin{equation}
\Omega ^{\alpha \beta }=\frac{\partial g^{\alpha }(\mathbf{\xi })}{\partial
\xi _{\beta }}-\frac{\partial g^{\beta }(\mathbf{\xi })}{\partial \xi
_{\alpha }},\qquad \alpha ,\beta =1,2,3  \label{b1.21}
\end{equation}
\begin{equation}
\delta \eta :\qquad j^{k}\partial _{k}S=0,  \label{b3.23}
\end{equation}
\begin{equation}
\delta S:\qquad j^{k}\partial _{k}\eta =-\rho \frac{\partial E}{\partial S},
\label{b3.24}
\end{equation}
\begin{equation}
\delta \mathbf{j}:\qquad \mathbf{v}\equiv \mathbf{j}/\rho =\nabla \varphi
+g^{\alpha }(\mathbf{\xi })\nabla \xi _{\alpha }-\eta \nabla S,
\label{b3.25}
\end{equation}
\begin{equation}
\delta \rho :\qquad -{\frac{\mathbf{j}^{2}}{2\rho ^{2}}}-\frac{\partial
(\rho E)}{\partial \rho }-\partial _{0}\varphi -g^{\alpha }(\mathbf{\xi }%
)\partial _{0}\xi _{\alpha }+\eta \partial _{0}S=0,  \label{b3.26}
\end{equation}
Deriving relations (\ref{b3.22}), (\ref{b3.24}), the continuity equation (%
\ref{b3.21}) is used. It is easy to see that (\ref{b3.22}) is equivalent to
the Lin constraints (\ref{b1.4}), provided 
\begin{equation}
\det \parallel \Omega ^{\alpha \beta }\parallel \neq 0  \label{b3.27}
\end{equation}

If condition (\ref{b3.27}) obtains, and equations (\ref{b1.4}) are
satisfied, the equations (\ref{b3.23}) and (\ref{b3.21}) can be integrated
in the form of (\ref{b1.9}) and (\ref{b1.10}) respectively. Equations (\ref
{b3.24}) and (\ref{b3.25}) are equivalent to (\ref{b1.13}) and (\ref{b1.11}%
). Finally, eliminating $\partial _{0}\xi _{\alpha }$ and $\partial _{0}S$
from (\ref{b3.26}) by means of (\ref{b1.4}) and (\ref{b3.23}), we obtain the
equation (\ref{b1.12}) and, hence, the system of dynamic equations (\ref
{b1.4}), (\ref{b1.12}), (\ref{b1.13}), where designations (\ref{b1.9})--(\ref
{b1.11}) are used.

The curtailed system (\ref{b1.1})--(\ref{b1.3}) can be obtained from
equations (\ref{b3.21})--(\ref{b3.26}) as follows. Equations (\ref{b3.21}), (%
\ref{b3.23}) coincide with (\ref{b1.1}), (\ref{b1.3}). For deriving (\ref
{b1.2}) let us note that the vorticity $\mathbf{\omega }\equiv \nabla \times 
\mathbf{v}$ and $\mathbf{v}\times \mathbf{\omega }$ are obtained from (\ref
{b3.25}) in the form 
\begin{equation}
\mathbf{\omega }=\mathbf{\nabla }\times \mathbf{v}={\frac{1}{2}}{\Omega }%
^{\alpha \beta }\mathbf{\nabla }\xi _{\beta }\times \mathbf{\nabla }\xi
_{\alpha }-\mathbf{\nabla }\eta \times \mathbf{\nabla }S  \label{b3.28}
\end{equation}
\begin{equation}
\mathbf{v}\times \mathbf{\omega }={\Omega }^{\alpha \beta }\mathbf{\nabla }%
\xi _{\beta }(\mathbf{v\nabla })\xi _{\alpha }+\mathbf{\nabla }S(\mathbf{%
v\nabla })\eta -\mathbf{\nabla }\eta (\mathbf{v\nabla })S  \label{b3.29}
\end{equation}

Let us form a difference between the time derivative of (\ref{b3.25}) and
the gradient of (\ref{b3.26}). Eliminating ${\Omega }^{\alpha \beta
}\partial _{0}\mathbf{\xi }_{\alpha }$, $\partial _{0}S$ and $\partial
_{0}\eta $ by means of equations (\ref{b3.22}), (\ref{b3.23}), (\ref{b3.24}%
), one obtains 
\[
\partial _{0}\mathbf{v}+\mathbf{\nabla }\frac{\mathbf{v}^{2}}{2}+\frac{%
\partial ^{2}(\rho E)}{\partial \rho ^{2}}\mathbf{\nabla }\rho +\frac{%
\partial ^{2}(\rho E)}{\partial \rho \partial S}\mathbf{\nabla }S-\rho \frac{%
\partial E}{\partial S}\mathbf{\nabla }S 
\]
\begin{equation}
-\Omega ^{\alpha \beta }\mathbf{\nabla }\xi _{\beta }(\mathbf{v\nabla })\xi
_{\alpha }+\mathbf{\nabla }\eta (\mathbf{v\nabla })S-\mathbf{\nabla }S(%
\mathbf{v\nabla })\eta =0  \label{b3.30}
\end{equation}
Using (\ref{b3.28}), (\ref{b3.29}), the expression (\ref{b3.30}) reduces to 
\begin{equation}
\partial _{0}\mathbf{v}+\mathbf{\nabla }{\frac{\mathbf{v}^{2}}{2}}+\frac{1}{%
\rho }\mathbf{\nabla }(\rho ^{2}\frac{\partial E}{\partial \rho })-\mathbf{v}%
\times (\mathbf{\nabla }\times \mathbf{v})=0  \label{b3.31}
\end{equation}
In virtue of the identity 
\begin{equation}
\mathbf{v}\times (\mathbf{\nabla }\times \mathbf{v})\equiv \mathbf{\nabla }{%
\frac{\mathbf{v}^{2}}{2}}-(\mathbf{v\nabla })\mathbf{v}  \label{b3.32}
\end{equation}
the last equation is equivalent to (\ref{b1.2}). Note that at derivation of
the curtailed system (\ref{b1.1}) -- (\ref{b1.3}) the condition (\ref{b3.27}%
) is not used, and the system (\ref{b1.1}) -- (\ref{b1.3}) is valid in any
case.

In the general case (\ref{b3.27}), differentiating equations (\ref{b3.25}), (%
\ref{b3.26}) and eliminating the variables $\varphi $, $\mathbf{\xi }$, $%
\eta $, we obtain the curtailed system (\ref{b1.1}) -- (\ref{b1.3}), whereas
the system (\ref{b1.4}), (\ref{b1.12}), (\ref{b1.13}) follows from the
system (\ref{b3.21}) -- (\ref{b3.26}) directly (i.e. without
differentiating). It means that the system (\ref{b1.4}), (\ref{b1.12}), (\ref
{b1.13}) is an integrated system, whereas the curtailed system (\ref{b1.1})
-- (\ref{b1.3}) is not, although formally they have the same order.

The action of the form (\ref{b3.20}), or close to this form was obtained by
some authors \cite{SW67,S88}, but the quantities $g^\alpha $, $\alpha =1,2,3$
are always considered as additional dependent variables (but not as
indefinite functions of $\mathbf{\xi }$ which can be expressed via initial
conditions). The action was not considered as a functional of fixed
indefinite functions $g^\alpha (\mathbf{\xi })$.

Thus, five equations (\ref{b1.4}), (\ref{b1.12}), (\ref{b1.13}) with $S$, $%
\rho $ and $\mathbf{v}$, defined respectively by (\ref{b1.9}), (\ref{b1.10})
and (\ref{b1.11}), constitute the fifth order system for five dependent
variables $\xi =\{\xi _{0},\mathbf{\xi }\}$, $\eta $. Equations (\ref{b1.1}%
), (\ref{b1.3}),(\ref{b1.4}), (\ref{b1.12}),(\ref{b1.13}) constitute the
seventh order system for seven variables $\rho $, $\mathbf{\xi }$, $\varphi $%
, $\eta $, $S$.

If $\Omega ^{\alpha \beta }\equiv 0$, it follows from (\ref{b1.21}) that 
\begin{equation}
g^{\alpha }\left( \mathbf{\xi }\right) =\frac{\partial }{\partial \xi
_{\alpha }}\Phi \left( \mathbf{\xi }\right)  \label{b3.27a}
\end{equation}
Then it follows from (\ref{b3.25}) and (\ref{b3.27a}) 
\begin{equation}
\mathbf{v}=\mathbf{j}/\rho =\nabla \left( \varphi +\Phi \left( \mathbf{\xi }%
\right) \right) -\eta \nabla S,  \label{b3.27b}
\end{equation}
In the case of isentropic flow $\left( \nabla S=0\right) $ the quantities $%
\Omega ^{\alpha \beta }$ coincide with vorticity, and the fluid flow is
irrotational, as it follows from (\ref{b3.27}). In this case, as well as at
fulfillment of (\ref{b3.27b}) the dynamic equations (\ref{b3.22}) are
satisfied due to relations $\Omega ^{\alpha \beta }\equiv 0$, and Lin
constraints (\ref{b1.4}) do not follow from dynamic equations (\ref{b3.22}).
In this partial case an addition of the Lin constraints (\ref{b1.4}) to the
curtailed system (\ref{b1.1}), (\ref{b1.3}) is not necessary.

The system of equations (\ref{b3.21}) -- (\ref{b3.26}), as well as the
system (\ref{b1.4}), (\ref{b1.9}) - (\ref{b1.13}) contain full information
on the fluid flow in the infinite space $V$. The system of equations (\ref
{b1.4}), (\ref{b1.9}) - (\ref{b1.13}) is a system of five partial
differential equations for five dynamic variables $\varphi $, $\eta $, $%
\mathbf{\xi }$, and one needs to give initial data for them. But the initial
values for variables $\varphi $, $\eta $, $\mathbf{\xi }$ can be given in
the universal form, which is the same for all fluid flows. For instance, one
can set 
\begin{equation}
\mathbf{\xi }\left( 0,\mathbf{x}\right) =\mathbf{\xi }_{\mathrm{in}}\left( 
\mathbf{x}\right) =\mathbf{x},\qquad \varphi \left( 0,\mathbf{x}\right)
=0,\qquad \eta \left( 0,\mathbf{x}\right) =0  \label{h3.28}
\end{equation}
Then according to (\ref{b3.25}) the initial values of variables $\rho $, $S$%
, $\mathbf{v}$ have the form 
\begin{eqnarray}
\mathbf{v}\left( 0,\mathbf{x}\right) &=&\mathbf{v}_{\mathrm{in}}\left( 
\mathbf{x}\right) =\mathbf{g}\left( \mathbf{\xi }\right) =\mathbf{g}\left( 
\mathbf{x}\right) ,  \label{h3.29} \\
\rho \left( 0,\mathbf{x}\right) &=&\rho _{\mathrm{in}}\left( \mathbf{\xi }%
\right) =\rho _{\mathrm{in}}\left( \mathbf{x}\right) ,\qquad S\left( 0,%
\mathbf{x}\right) =S_{\mathrm{in}}\left( \mathbf{\xi }\right) =S_{\mathrm{in}%
}\left( \mathbf{x}\right)  \label{h3.30}
\end{eqnarray}
where $\rho _{\mathrm{in}}\left( \mathbf{x}\right) $, $S_{\mathrm{in}}\left( 
\mathbf{x}\right) $, $\mathbf{v}_{\mathrm{in}}\left( \mathbf{x}\right) $ are
given initial values which determine the fluid flow. Variables $\mathbf{\xi }
$ label the fluid particles, and a choice of the form of labeling is
unessential.

Let now the form of $\varphi \left( 0,\mathbf{x}\right) $ and $\eta \left( 0,%
\mathbf{x}\right) $ be changed and, we have $\varphi \left( 0,\mathbf{x}%
\right) =\varphi _{\mathrm{in}}\left( \mathbf{x}\right) $, $\eta \left( 0,%
\mathbf{x}\right) =\eta _{\mathrm{in}}\left( \mathbf{x}\right) $, where $%
\varphi _{\mathrm{in}}\left( \mathbf{x}\right) $, and $\eta _{\mathrm{in}%
}\left( \mathbf{x}\right) $ are some given functions. This replacement can
be compensated by the change of $\mathbf{g}\left( \mathbf{x}\right) $%
\[
\mathbf{g}\left( \mathbf{x}\right) \rightarrow \mathbf{v}_{\mathrm{in}%
}\left( \mathbf{x}\right) -\mathbf{\nabla }\varphi _{\mathrm{in}}\left( 
\mathbf{x}\right) +\eta _{\mathrm{in}}\left( \mathbf{x}\right) \mathbf{%
\nabla }S_{\mathrm{in}}\left( \mathbf{x}\right) 
\]
in such a way that the initial values of $\rho $, $S$, $\mathbf{v}$ remain
the same. All this mean that dynamic equations (\ref{b1.4}), (\ref{b1.9}) - (%
\ref{b1.13}) contain complete (full) information on the fluid flow, because
the choice of initial values for variables $\varphi $, $\eta $, $\mathbf{\xi 
}$ is unimportant for calculation of the fluid flow in the infinite space $V$%
.

This is valid also for the fluid flow in the space region $x^{3}\geq 0$, as
it is shown in \cite{R999}. In this case the boundary conditions for $%
\mathbf{v}$ at the boundary $x^{3}=0$ are expressed via the arbitrary
functions $\mathbf{g}\left( \mathbf{\xi }\right) $, $\mathbf{\xi }=\mathbf{%
x\in }\left\{ \mathbf{x}|x^{3}<0\right\} $, whereas the initial condition
are expressed via the arbitrary functions $\mathbf{g}\left( \mathbf{\xi }%
\right) $, $\mathbf{\xi }=\mathbf{x\in }\left\{ \mathbf{x}|x^{3}\geq
0\right\} $. Apparently, functions $\mathbf{g}\left( \mathbf{x}\right) $
determine the initial and boundary conditions for the velocity $\mathbf{v}$
also in the case of the fluid flow in any finite volume $V$, although it is
not yet proved.

Thus, the information, which is essential for the fluid flow determination,
is described by functions $\rho _{\mathrm{in}}\left( \mathbf{x}\right) $, $%
S_{\mathrm{in}}\left( \mathbf{\xi }\right) $, $\mathbf{g}$ . This
information is introduced in dynamic equations in the form of arbitrary
functions. Unessential information concerning the methods of the fluid
description, is given by initial and boundary conditions for variables $%
\varphi $, $\eta $, $\mathbf{\xi }$. The variables $\mathbf{\xi }$, $\varphi 
$, $\eta $ are auxiliary variables which represent a method of the fluid
flow description. The Clebsch potentials $\mathbf{\xi }$ label the fluid
particles. At the same time the variables $\mathbf{\xi }$ describe
displacement of the fluid along their trajectories, and this description
does not depend on the method of labeling. Variables $\varphi $, $\eta $
describe separation of the velocity field $\mathbf{v}$ into parts. The form
of this separation is inessential. The auxiliary variables $\mathbf{\xi }$, $%
\varphi $, $\eta $ are described by partial differential equations, whereas
the fluid flow in itself is described by finite relations containing
arbitrary functions $\rho _{\mathrm{in}}\left( \mathbf{\xi }\right) $, $S_{%
\mathrm{in}}\left( \mathbf{\xi }\right) $, $\mathbf{g}\left( \mathbf{\xi }%
\right) $ (at fixed variables $\varphi $, $\eta $, $\mathbf{\xi }$, given as
functions of $t,\mathbf{x}$).

Situation is rather unexpected and unusual. One can obtain a result of
calculation of the fluid flow, but it refers to the method of description
(variables $\mathbf{\xi }$, $\varphi $, $\eta $), and this method of
description is determined by partial differential equations. It is not clear
how to resolve and to use this situation effectively.

\section{Variational principle for incompressible fluid}

Inviscid incompressible fluid of constant density is a special case of ideal
fluid, when one may set $\rho =\rho _{0}=$const. In this case the continuity
equation (\ref{b3.21}), containing time derivative of density $\rho $ and
determining time evolution of $\rho $, becomes 
\begin{equation}
\mathbf{\nabla v=0}  \label{b6.1a}
\end{equation}
Equation (\ref{b6.1a}) does not contain time derivatives at all. It is
rather a constraint imposed on initial values of velocity $\mathbf{v}$, than
a dynamic equation, describing evolution of one of dynamic variables.
Formally it means that the system of hydrodynamic equations ceases to be
hyperbolic and becomes elliptic. This circumstance changes the statement of
the fluid flow problem.

Let us set $\rho =\rho _{0}=\mathrm{const}$ in the action (\ref{b3.20}) and
introduce new variables 
\begin{equation}
\mathbf{v}=\mathbf{j}/\rho _{0},\qquad \rho _{0}=\mathrm{const}  \label{b6.1}
\end{equation}
It is easy to verify that $\eta =\eta (\mathbf{\xi })$, $S=S_{0}(\mathbf{\xi 
})$, and the last term of (\ref{b3.20}) can be incorporated with the term $%
j^{k}g^{\alpha }(\mathbf{\xi })\partial _{k}\mathbf{\xi }_{\alpha }$. Thus,
the action for the incompressible fluid looks as follows 
\begin{equation}
\mathcal{A}_{\mathrm{E}}[\mathbf{v},\mathbf{\xi },\varphi ]=\rho _{0}\int \{{%
\frac{\mathbf{v}^{2}}{2}}-\mathbf{v\nabla }\varphi -g^{\alpha }(\mathbf{\xi }%
)\partial _{0}\xi _{\alpha }-g^{\alpha }(\mathbf{\xi })\mathbf{v\nabla }\xi
_{\alpha }\}dtd\mathbf{x},  \label{b6.2}
\end{equation}
where $g^{\alpha }(\mathbf{\xi })$ are arbitrary fixed functions of $\mathbf{%
\xi }$.

Variation with respect to $\mathbf{v}$, $\mathbf{\xi }$, $\varphi $ gives 
\begin{equation}
\delta \mathbf{v}:\qquad \mathbf{v}=\mathbf{\nabla }\varphi +g^{\alpha }(%
\mathbf{\xi })\mathbf{\nabla }\xi _{\alpha }  \label{b6.3}
\end{equation}
\begin{equation}
\rho _{0}^{-1}{\frac{\delta \mathcal{A}_{\mathrm{E}}}{\delta \xi _{\alpha }}}%
=\Omega ^{\alpha \beta }(\partial _{0}\xi _{\beta }+\mathbf{v\nabla }\xi
_{\beta })=0,\qquad \alpha =1,2,3  \label{b6.4}
\end{equation}
\begin{equation}
\rho _{0}^{-1}{\frac{\delta \mathcal{A}_{\mathrm{E}}}{\delta \varphi }}=%
\mathbf{\nabla v}=0  \label{b6.5}
\end{equation}

In the general case the condition (\ref{b3.27}) is satisfied, and the
multiplier $\Omega ^{\alpha \beta }$ in (\ref{b6.4}) may be omitted.

Substituting (\ref{b6.3}) into (\ref{b6.4}) and (\ref{b6.5}), one obtains 
\begin{equation}
\Omega ^{\alpha \beta }\left( \partial _{0}\xi _{\beta }+\left( \mathbf{%
\nabla }\varphi +g^{\gamma }(\mathbf{\xi })\mathbf{\nabla }\xi _{\gamma
}\right) \mathbf{\nabla }\xi _{\beta }\right) =0,\qquad \alpha =1,2,3
\label{b6.6}
\end{equation}
\begin{equation}
\mathbf{\nabla }^{2}\varphi +g^{\alpha ,\beta }(\mathbf{\xi })\mathbf{\nabla 
}\xi _{\beta }\mathbf{\nabla }\xi _{\alpha }+g^{\alpha }(\mathbf{\xi })%
\mathbf{\nabla }^{2}\xi _{\alpha }=0,\qquad g^{\alpha ,\beta }\equiv \frac{%
\partial g^{\alpha }}{\partial \xi _{\beta }}  \label{b6.7}
\end{equation}
The dynamic equation for $\varphi $ does not contain temporal derivative. If 
$\Omega ^{\alpha \beta }\equiv 0$, the fluid flow is irrotational and
dynamic equations (\ref{b6.6}) are fulfilled independently of the Lin
constraints, which have the form 
\begin{equation}
\partial _{0}\xi _{\beta }+\left( \mathbf{\nabla }\varphi +g^{\gamma }(%
\mathbf{\xi })\mathbf{\nabla }\xi _{\gamma }\right) \mathbf{\nabla }\xi
_{\beta }=0,\qquad \alpha =1,2,3  \label{b6.7a}
\end{equation}
Lin constraints (\ref{b6.7a}) are not dynamic equations in this case.

Conventional hydrodynamic equations for the incompressible fluid 
\begin{equation}
\mathbf{\nabla v}=0,\qquad \partial _{0}\mathbf{v}+(\mathbf{v\nabla })%
\mathbf{v}=-\frac{\mathbf{\nabla }p}{\rho _{0}}  \label{b6.8}
\end{equation}
are obtained from relations (\ref{b6.3})-(\ref{b6.5}). Differentiating (\ref
{b6.3}) with respect to $t$, we obtain 
\begin{equation}
\partial _{0}\mathbf{v}=\mathbf{\nabla }[\partial _{0}\varphi +g^{\alpha }(%
\mathbf{\xi })\partial _{0}\xi _{\alpha }]-\Omega ^{\alpha \beta }\partial
_{0}\xi _{\beta }\mathbf{\nabla }\xi _{\alpha }  \label{b6.9}
\end{equation}
where $\Omega ^{\alpha \beta }$ is defined by (\ref{b1.21}). It follows from
(\ref{b6.3}) 
\begin{equation}
\mathbf{v\times }(\mathbf{\nabla }\times \mathbf{v})=\Omega ^{\alpha \beta }(%
\mathbf{\xi })\mathbf{\nabla }\xi _{\beta }(\mathbf{v\nabla })\xi _{\alpha }.
\label{b6.10}
\end{equation}
In virtue of (\ref{b6.4}) the last term in rhs of (\ref{b6.9}) coincides
with rhs of (\ref{b6.10}). Then using the identity (\ref{b3.32}) one obtains 
\begin{equation}
\partial _{0}\mathbf{v}+(\mathbf{v\nabla })\mathbf{v}=\mathbf{\nabla }\left(
\partial _{0}\varphi +g^{\alpha }(\mathbf{\xi })\partial _{0}\xi _{\alpha }+{%
\frac{1}{2}}\mathbf{v}^{2}\right)  \label{b6.11}
\end{equation}
The equation (\ref{b6.11}) coincides with the second equation (\ref{b6.8}),
provided we use designation 
\begin{equation}
\frac{p}{\rho _{0}}=\frac{p_{0}}{\rho _{0}}-{\frac{1}{2}}\mathbf{v}%
^{2}-\partial _{0}\varphi -g^{\alpha }(\mathbf{\xi })\partial _{0}\xi
_{\alpha },\qquad p_{0}=\text{const}  \label{b6.12}
\end{equation}
Here the pressure $p$ is determined after solution of the system of
hydrodynamic equations (\ref{b6.3})-(\ref{b6.5}), or (\ref{b6.8}).

Let us stress that the conventional form (\ref{b6.8}) of hydrodynamic
equations is obtained from the hydrodynamic equations (\ref{b6.3})-(\ref
{b6.5}) \textit{by means of differentiation}. It means that the form of
hydrodynamic equations (\ref{b6.3})-(\ref{b6.5}) is a \textit{result of
integration} of hydrodynamic equations (\ref{b6.8}) together with the Lin
constraints (\ref{b1.4}). It is interesting also that the system of
equations (\ref{b6.3})-(\ref{b6.5}) contains time derivatives only in
dynamic equations (\ref{b6.4}).

\section{Two-dimensional flow of incompressible fluid}

We have mentioned in introduction that statement of the problem of stream
flow is different for the irrotational and rotational cases. This difference
appears only after integration. The statement of the problem is different
not only for two-dimensional flow of incompressible fluid. It is different
for any inviscid fluid, and the source of this difference lies in the
dynamic equations (\ref{b3.22}) (or (\ref{b6.4})) which exclude the degrees
of freedom connected with rotation in the case $\Omega ^{\alpha \beta
}\equiv 0$. The method used for derivation of (\ref{h1.1a}) cannot be used
in general case. So in the following we present the method in a form which
can be expanded to any ideal fluid. The method is based on a use of the
generalized stream function (mutual application of the Jacobian technique
and description in terms of potentials). To avoid technical complexities we
apply this method to the case of two-dimensional flow of incompressible
fluid, when the generalized stream function has only one component.

Equations (\ref{h1.1}) and (\ref{h1.1a}) are different. The first equation
is linear, whereas the second one is quasilinear. The first one can be
solved rather easily, whereas the second one can be solved only by means of
an iteration procedure. Difference in complexity of equations (\ref{h1.1})
and (\ref{h1.1a}) is technical, whereas the difference in statement of the
problem is conceptual. The idea of our investigation is simple. Introducing
the generalized stream function $\xi _{2}$ we solve equations (\ref{b6.4}), (%
\ref{b6.5}) for any $\xi _{2}$. Then from (\ref{b6.3}) we obtain dynamic
equation for determination of $\xi _{2}$ and necessary boundary conditions
for this equation.

Although it is possible to deal with equations (\ref{b6.3}) -- (\ref{b6.5})
for incompressible fluid, we prefer to consider dynamic equations for
slightly compressible fluid, whose internal energy has the form 
\begin{equation}
E\left( \rho \right) =E_{0}\left( \frac{\rho }{\rho _{0}}\right)
^{1/\varepsilon },\qquad E_{0},\rho _{0}=\text{const, \qquad }\varepsilon
\ll 1  \label{h2.1}
\end{equation}
The incompressible fluid appears in the limit $\varepsilon \rightarrow 0$.
We consider dynamic equations obtained from the action (\ref{b3.20}) with
the entropy $S=$ const and $\eta =0$. These equations have the form

\begin{eqnarray}
\partial _{k}j^{k} &=&0,\qquad j^{0}\equiv \rho  \label{h2.3} \\
\Omega ^{\alpha \beta }j^{k}\xi _{\beta ,k} &=&0,\qquad \alpha =1,2
\label{h2.4}
\end{eqnarray}
where $\Omega ^{\alpha \beta }$ is defined by the relation (\ref{b1.21}). 
\begin{eqnarray}
\frac{j^{\mu }}{\rho } &=&\partial _{\mu }\varphi +g^{\alpha }(\mathbf{\xi }%
)\xi _{\alpha ,\mu },\qquad \mu =1,2,  \label{h2.5} \\
\xi _{\alpha ,\mu } &\equiv &\partial _{\mu }\xi _{\alpha },\qquad \mu
,\alpha =1,2  \nonumber
\end{eqnarray}
\begin{equation}
\frac{\partial \left( \rho E\right) }{\partial \rho }=-\frac{\mathbf{j}^{2}}{%
2\rho ^{2}}-\partial _{0}\varphi +g^{\alpha }(\mathbf{\xi })\xi _{\alpha ,0}
\label{h2.6}
\end{equation}
Here and in what follows a summation is produced over repeating Greek
indices $(1,2)$ and over repeating Latin indices ($0-2$). We use
designations $(x^{0},x^{1},x^{2})=(t,x,y)$. Comma before index $k$ denotes
differentiation with respect to $x^{k}$, for instance, $\xi _{\alpha
,k}\equiv \partial _{k}\xi _{\alpha }$.

Let us set 
\begin{equation}
J=\frac{\partial \left( \xi _{0},\xi _{1},\xi _{2}\right) }{\partial \left(
x^{0},x^{1},x^{2}\right) },\qquad j^{k}=\frac{\partial J}{\partial \xi _{0,k}%
}=\frac{\partial \left( x^{k},\xi _{1},\xi _{2}\right) }{\partial \left(
x^{0},x^{1},x^{2}\right) },\qquad k=0,1,2,  \label{h2.7}
\end{equation}
where $\xi _{1},\xi _{2}$ are some functions of $\left( t,x,y\right) $.
Equations (\ref{h2.3}), (\ref{h2.4}) are satisfied by (\ref{h2.7}) for any
functions $\xi _{1},\xi _{2}$, because of the identities 
\begin{equation}
\partial _{k}\frac{\partial J}{\partial \xi _{0,k}}\equiv 0,\qquad \frac{%
\partial J}{\partial \xi _{0,k}}\xi _{\alpha ,k}\equiv 0,\qquad \alpha =1,2
\label{h2.8}
\end{equation}
Substituting (\ref{h2.1}) in (\ref{h2.6}), we obtain 
\begin{equation}
\left( E_{0}\right) ^{\varepsilon }\left( 1+\frac{1}{\varepsilon }\right)
^{\varepsilon }\left( \frac{\rho }{\rho _{0}}\right) =\left( -{\frac{\mathbf{%
j}^{2}}{2\rho ^{2}}}-\partial _{0}\varphi +g^{\alpha }(\mathbf{\xi })\xi
_{\alpha ,0}\right) ^{\varepsilon }  \label{h2.9}
\end{equation}
In the limit $\varepsilon \rightarrow 0$ equation (\ref{h2.9}) becomes 
\begin{equation}
\frac{\rho }{\rho _{0}}=\frac{1}{\rho _{0}}\frac{\partial \left( \xi
_{1},\xi _{2}\right) }{\partial \left( x^{1},x^{2}\right) }=1  \label{h2.10}
\end{equation}

In the case of stationary flow we can assume that $\xi _{2}$ does not depend
on the time $t$ 
\begin{equation}
\xi _{1}=\xi _{1}\left( t,x,y\right) ,\qquad \xi _{2}=\xi _{2}\left(
x,y\right)   \label{h2.11}
\end{equation}
and 
\begin{equation}
\xi _{2,0}\equiv 0.  \label{h2.16}
\end{equation}

Equation (\ref{h2.10}) takes the form 
\begin{equation}
\xi _{2,2}\xi _{1,1}-\xi _{2,1}\xi _{1,2}-\rho _{0}=0.  \label{h2.17}
\end{equation}
Let us consider equation (\ref{h2.17}) as an equation for dependent variable 
$\xi _{1}$ with a given function $\xi _{2}=\xi _{2}\left( x,y\right) $. The
first order partial differential equation is equivalent to the system of
ordinary equations 
\begin{equation}
\frac{dx}{\xi _{2,2}}=\frac{dy}{-\xi _{2,1}},\qquad \frac{dx}{\xi _{2,2}}=%
\frac{d\xi _{1}}{\rho _{0}}.  \label{h2.18}
\end{equation}

It is easy to see that 
\begin{equation}
C_{2}=\xi _{2}\left( x,y\right)  \label{h2.19}
\end{equation}
is an integral of the first equation (\ref{h2.18}). Resolving equation (\ref
{h2.19}) with respect to $y$, one obtains 
\begin{equation}
y=F_{2}\left( x,C_{2}\right) ,\qquad \xi _{2}\left( x,F_{2}\left(
x,C_{2}\right) \right) =C_{2}  \label{h2.20}
\end{equation}
Substituting (\ref{h2.20}) in the second equation (\ref{h2.18}), one can
integrate it in the form 
\begin{equation}
\xi _{1}=\rho _{0}\left[ C_{1}\left( t,C_{2}\right) +\int\limits_{-L}^{x}%
\frac{dx}{\xi _{2,2}\left( x,F_{2}\left( x,C_{2}\right) \right) }\right]
_{C_{2}=\xi _{2}\left( x,y\right) }  \label{h2.21}
\end{equation}
where $C_{1}$ is an arbitrary function of arguments $\left( t,C_{2}\right) $%
. Here $x=-L$ is the inflow surface $\mathcal{S}_{\mathrm{in}}$, where the
inflow conditions are given. The inflow surface is placed in the fluid, and
all dynamic equations must be satisfied on $\mathcal{S}_{\mathrm{in}}$.

Differentiating the second equation (\ref{h2.20}) with respect to $C_{2}$,
we obtain 
\begin{equation}
\xi _{2,2}\left( x,F_{2}\left( x,C_{2}\right) \right) F_{2;2}\left(
x,C_{2}\right) =1  \label{h2.21a}
\end{equation}
Here and in what follows the symbol ';' denotes differentiation with respect
to one of arguments of the function. Index after symbol ';' denotes the
ordinal number of the argument. By means of (\ref{h2.21a}) integral in (\ref
{h2.21}) may be written in the form 
\begin{equation}
\int\limits_{-L}^{x}\frac{dx}{\xi _{2,2}\left( x,F_{2}\left( x,C_{2}\right)
\right) }=\frac{\partial }{\partial C_{2}}\int\limits_{-L}^{x}F_{2}\left(
x,C_{2}\right) dx  \label{h2.21b}
\end{equation}
The relation (\ref{h2.21}) takes the form 
\begin{equation}
\xi _{1}=\rho _{0}\left[ C_{1}\left( t,C_{2}\right) +\frac{\partial }{%
\partial C_{2}}\int\limits_{-L}^{x}F_{2}\left( x,C_{2}\right) dx\right]
_{C_{2}=\xi _{2}\left( x,y\right) }  \label{h2.21c}
\end{equation}

Differentiating (\ref{h2.21}) with respect to $t$, we obtain 
\begin{equation}
\xi _{1,0}=\rho _{0}\left[ \frac{\partial C_{1}\left( t,C_{2}\right) }{%
\partial t}\right] _{C_{2}=\xi _{2}\left( x,y\right) }=\rho
_{0}C_{1;1}\left( t,\xi _{2}\right)  \label{h2.22a}
\end{equation}

According to (\ref{h2.11}) and (\ref{h2.7}) we have 
\begin{equation}
j^{1}=-\frac{\partial \left( \xi _{1},\xi _{2}\right) }{\partial \left(
x^{0},x^{2}\right) }=-\xi _{1,0}\xi _{2,2},\qquad j^{2}=\frac{\partial
\left( \xi _{1},\xi _{2}\right) }{\partial \left( x^{0},x^{1}\right) }=\xi
_{1,0}\xi _{2,1}  \label{h2.25}
\end{equation}
and equations (\ref{h2.5}) take the form 
\begin{equation}
u=v^{1}=-\frac{\xi _{1,0}\xi _{2,2}}{\rho _{0}}=\partial _{1}\varphi
+g^{\alpha }(\mathbf{\xi })\xi _{\alpha ,1}  \label{h2.26}
\end{equation}
\begin{equation}
v=v^{2}=\frac{\xi _{1,0}\xi _{2,1}}{\rho _{0}}=\partial _{2}\varphi
+g^{\alpha }(\mathbf{\xi })\xi _{\alpha ,2}  \label{h2.27}
\end{equation}

The compatibility condition of equations (\ref{h2.26}) and (\ref{h2.27}) has
the form 
\begin{equation}
-\partial _{2}\frac{\xi _{1,0}\xi _{2,2}}{\rho _{0}}-\partial _{1}\frac{\xi
_{1,0}\xi _{2,1}}{\rho _{0}}=\left( g^{\alpha ,\beta }(\mathbf{\xi }%
)-g^{\beta ,\alpha }(\mathbf{\xi })\right) \xi _{\alpha ,1}\xi _{\beta
,2},\qquad g^{\alpha ,\beta }\equiv \frac{\partial g^{\alpha }}{\partial \xi
_{\beta }}  \label{h2.28}
\end{equation}
Substituting (\ref{h2.22a}) and using designation (\ref{b1.21}) we obtain

\begin{equation}
\psi _{,11}+\psi _{,22}=-\rho _{0}\Omega ^{12}(\xi _{1},\xi _{2}),
\label{h2.32}
\end{equation}
where the stream function $\psi $ is defined by the relation 
\begin{equation}
\psi =C_{0}\left( t,\xi _{2}\right) =\int C_{1;1}\left( t,\xi _{2}\right)
d\xi _{2}  \label{hh2.27}
\end{equation}
Relation (\ref{hh2.27}) is chosen in such a way that equations (\ref{h2.26}%
), (\ref{h2.27}) coincide with (\ref{h1.2b}). Let us resolve relation (\ref
{hh2.27}) with respect to $\xi _{2}$ in the form $\xi _{2}=Q\left( t,\psi
\right) $, where $Q\left( t,\psi \right) $ is the functional inverse of $%
C_{0}\left( t,\xi _{2}\right) $ 
\begin{equation}
C_{0}\left( t,Q\left( t,\psi \right) \right) =\psi  \label{hh2.28}
\end{equation}
Substituting $\xi _{2}$ in (\ref{h2.21}), we obtain Clebsch potentials $\xi
_{1}$ and $\xi _{2}$ expressed via $\psi $ by means of relations 
\begin{equation}
\xi _{2}=Q\left( t,\psi \right) ,\qquad \xi _{1}=\xi _{1}\left( t,x,\psi
\right) =\rho _{0}\left[ C_{1}\left( t,C_{2}\right) +\frac{\partial }{%
\partial C_{2}}\int\limits_{-L}^{x}F_{2}\left( x,C_{2}\right) dx\right]
_{C_{2}=Q\left( t,\psi \right) }  \label{h2.33}
\end{equation}
According to (\ref{h2.25}), (\ref{h2.22a}) and (\ref{hh2.27}) we obtain 
\begin{equation}
j^{1}=-\rho _{0}\psi _{,2},\qquad j^{2}=\rho _{0}\psi _{,1}  \label{h2.34}
\end{equation}
Dynamic equation (\ref{h2.32}) for the stream function $\psi \ $ is the
second order partial differential equation of elliptic type. It contains
indefinite functions $\Omega ^{12}(\xi _{1},\xi _{2}),$ $C_{0}\left( t,\xi
_{2}\right) $ which are determined from the inflow conditions. The time $t$
appears in dynamic equation only as a parameter describing evolution of the
Clebsch potential $\xi _{1}$. The dynamic equation (\ref{h2.32}) is
applicable for arbitrary two-dimensional flow of ideal incompressible fluid.

\textit{Remark. }Equation (\ref{h2.32}) can be obtained by direct
integration of Euler equations for two-dimensional flow of incompressible
fluid. Indeed, eliminating pressure from equations (\ref{b6.8}) for
two-dimensional flow, we obtain 
\begin{equation}
u_{,1}+v_{,2}=0,\qquad \partial _{2}\left( u_{,0}+uu_{,1}+vu_{,2}\right)
=\partial _{1}\left( v_{,0}+uv_{,1}+vv_{,2}\right)  \label{h2.35}
\end{equation}
where $u$ and $v$ are velocity components along $x$-axis and $y$-axis
respectively. Introducing the stream function $\psi $%
\begin{equation}
u=-\psi _{,2},\qquad v=\psi _{,1}  \label{h2.36}
\end{equation}
we satisfy the first equation (\ref{h2.35}) identically. The second equation
(\ref{h2.35}) takes the form 
\begin{equation}
\partial _{0}\omega -\frac{\partial \left( \omega ,\psi \right) }{\partial
\left( x,y\right) }=0,\qquad \omega \equiv \psi _{,11}+\psi _{,22}
\label{h2.37}
\end{equation}
In the stationary case, the result of integration is considered in
introduction.

In the case of arbitrary flow the equation (\ref{h2.37}) is reduced to the
form 
\begin{equation}
\frac{d\omega }{dt}=\partial _{0}\omega +u\partial _{1}\omega +v\partial
_{2}\omega =0  \label{h2.39}
\end{equation}
Solution of (\ref{h2.39}) can be presented in the form 
\begin{equation}
\omega =\psi _{,11}+\psi _{,22}=\Omega \left( \xi _{1},\xi _{2}\right)
\label{h2.40}
\end{equation}
where $\xi _{1}$ and $\xi _{2}$ are two independent solutions of equation (%
\ref{h2.39}), i.e. $\xi _{1}$ and $\xi _{2}$ are Clebsch potentials,
satisfying equations (\ref{b1.4}). Thus, the problem of the Euler system
integration appears to be connected with consideration of the Lin
constraints. The statement that we can investigate the fluid motion,
considering only the Euler system of dynamic equations and ignoring
completely Lin constraints, is valid only in the case $\Omega \equiv 0$. In
the general case it is only an illusion. In reality the Lin constraints are
hidden inside the Euler system and appear after integration.

In the simplest case, when all fluid particles cross boundary, and there are
no closed stream lines, the function $\Omega \left( \xi _{1},\xi _{2}\right) 
$ is determined from the inflow conditions on the boundary of the considered
flow. If the flow contains fluid particles which move all the time inside
the considered volume and do not cross its boundary, the function $\Omega
\left( \xi _{1},\xi _{2}\right) $ is to be given for these particles also,
although such an assignment is impossible via boundary conditions. Here we
shall consider only the simplest case of rotational flow, when there are no
closed stream lines. The case with closed stream lines of rotational flow
inside the irrotational flow is considered in the paper \cite{EFM01}. In
this paper the function $\Omega \left( \xi _{2}\right) =f\left( \psi \right) 
$ is given on the closed stream lines without a reference to boundary
conditions.

In the considered case the dynamic equation (\ref{h2.40}) appears to contain
information on boundary conditions. It means that, studying rotational
nonstationary flows, a separate investigation of dynamic equations and
boundary conditions becomes impossible. Appearance of several indefinite
functions of $\psi $ in dynamic equations of type (\ref{h1.1a}) for
stationary rotational flow is well known (see, for instance, survey \cite
{L72}). Unfortunately, an investigation of the origin of these functions is
sometimes absent.

Obtaining the basic equation (\ref{h2.32}), we used properties of Clebsch
potentials, described by equations (\ref{h2.5}). These relations contain
arbitrary functions $g^{\alpha }\left( \mathbf{\xi }\right) $, which
describe influence of boundary conditions on the form of dynamic equations.
Indefinite functions $g^{\alpha }\left( \mathbf{\xi }\right) $ describe
general properties of ideal fluid, which could be applied not only for
investigation of incompressible fluid. Information contained in three
functions $g^{a}\left( \mathbf{\xi }\right) $ of three arguments $\xi
_{1},\xi _{2},\xi _{3}$ is much more, than information contained in one
function $\Omega \left( \xi _{1},\xi _{2}\right) $ of two arguments. Maybe,
not all this information is essential, but some part of this information is
essential for sure. It is the reason why we investigate dynamic equations (%
\ref{b6.3}) --(\ref{b6.5}), but not Euler system (\ref{h2.35}), which
contain this information only in implicit form.

The last shorter derivation (\ref{h2.35}) -- (\ref{h2.40}) of equation (\ref
{h2.40}) ignores general relations (\ref{h2.4}), (\ref{h2.5}), but
nevertheless finally it leads to (\ref{h2.40}), i.e. to a description in
terms of Clebsch potentials.

\section{Inflow conditions for stationary flow}

Let us suppose that the stationary inflow conditions are given at the plane $%
\mathcal{S}_{\mathrm{in}}:$ $x=-L=$const. (We consider the case, when there
are no closed stream lines inside the considered volume.) The inflow surface 
$\mathcal{S}_{\mathrm{in}}$ lies in the fluid completely, and all dynamic
equations are satisfied on $\mathcal{S}_{\mathrm{in}}$. In the stationary
case the stream function $\psi $ does not depend on time $t$. One may set in
(\ref{hh2.27}) $C_{0}\left( t,\xi _{2}\right) =C_{0}\left( \xi _{2}\right) $%
. On the inflow surface we choose the function $\xi _{2}$ in the form 
\begin{equation}
\xi _{2}\left( -L,y\right) =y  \label{h3.0}
\end{equation}
According to (\ref{h2.25}) (\ref{h2.22a}) the inflow conditions have the
form 
\begin{eqnarray}
u\left( -L,y\right) &=&-C_{1;1}\left( -L,y\right) =V_{x}\left( y\right)
\label{h3.1} \\
\frac{\partial v}{\partial x}\left( -L,y\right) &=&\frac{1}{\rho _{0}}\left[
\partial _{1}\left( \xi _{1,0}\left( t,y\right) \xi _{2,1}\left( x,y\right)
\right) \right] _{x=-L}=W\left( y\right)  \label{h3.2}
\end{eqnarray}
where $u$ and $v$ are velocity components along $x$-axis and $y$-axis
respectively. $V_{x}\left( y\right) $ and $W\left( y\right) $ are supposed
to be known functions determining inflow conditions of the fluid. We obtain
from (\ref{h3.1}) and (\ref{hh2.27}) 
\begin{equation}
C_{1}\left( t,y\right) =-\int\limits_{0}^{t}V_{x}\left( y\right) dt\equiv
S\left( y\right) t,\qquad C_{0}\left( y\right) =-\int V_{x}\left( y\right) dy
\label{h3.3}
\end{equation}

The lhs of (\ref{h2.28}) at $x=-L$ is equal to 
\begin{equation}
\frac{\partial }{\partial y}u\left( -L,y\right) -\frac{\partial v}{\partial x%
}\left( -L,y\right) =\frac{\partial }{\partial y}V_{x}\left( y\right)
-W\left( y\right) ,  \label{h3.4}
\end{equation}
whereas rhs of (\ref{h2.28}) at $x=-L$ is equal to $\rho _{0}\Omega
^{12}(\xi _{1}\left( t,-L,y\right) ,y)$. Then using representation (\ref
{h2.21}), we obtain from (\ref{h2.28}), (\ref{h2.33}) at $x=-L$ 
\begin{equation}
\rho _{0}\Omega ^{12}(\rho _{0}C_{1}\left( t,y\right) ,y)=\frac{\partial }{%
\partial y}V_{x}\left( y\right) -W\left( y\right) \equiv U\left( y\right) \ 
\label{h3.5}
\end{equation}
where $U\left( y\right) $ is known function of variables $y$. As far as rhs
of (\ref{h3.5}) does not depend on $t$, lhs of (\ref{h3.5}) does not depend
on $C_{1}$, and $\Omega ^{12}$ depends only on $\xi _{2}$ 
\begin{equation}
\Omega ^{12}(\xi _{1},\xi _{2})=\Omega ^{12}(0,\xi _{2})=\frac{1}{\rho _{0}}%
U\left( \xi _{2}\right) =\frac{1}{\rho _{0}}\left( \frac{\partial }{\partial
\xi _{2}}V_{x}\left( \xi _{2}\right) -W\left( \xi _{2}\right) \right)
\label{h3.9}
\end{equation}

Thus, inflow conditions (\ref{h3.1}), (\ref{h3.2}) admit one to determine
indefinite functions $\Omega ^{12}$ and $C_{1}$. It is interesting, that
inflow conditions are given at the inflow surface $\mathcal{S}_{\mathrm{in}%
}:\;x=-L $ by the velocity component $u$ and by derivative $\partial
v/\partial x$ (normal to $\mathcal{S}_{\mathrm{in}}$) of the component $v$
(but not by the component $v$ itself). Information, giving by the velocity
component $v$ on the inflow surface $\mathcal{S}_{\mathrm{in}}$, is
insufficient for determination of indefinite functions $\Omega ^{12}$ and $%
C_{1}$. The derivative $\partial v/\partial x$ together with $\partial
u/\partial y$ determine vorticity. The derivative $\partial u/\partial y$
can be calculated, if the velocity $u$ is given on the inflow surface,
whereas $\partial v/\partial x$ cannot. To determine vorticity, we need to
give $\partial v/\partial x$ on the inflow surface, whereas the tangent
component $v$ itself appears to be unnecessary for the vorticity
determination.

\section{Problem of stationary flow over obstacle}

Let us consider a flow of incompressible fluid inside the space region $V$
bounded by two walls $\mathcal{S}_{1}:y=-R=$const$,$ $\;\;\mathcal{S}%
_{2}:\;y=R$,\ by inflow surface $\mathcal{S}_{\mathrm{in}}:$ \ $x=-L=$const,
by outflow surface $\mathcal{S}_{\mathrm{out}}:\ x=L$, and by the surface $%
\mathcal{S}_{\mathrm{ob}}$ of some obstacle, placed in vicinity of the
coordinate origin. The shape of the obstacle is determined by the relation 
\begin{equation}
\mathcal{S}_{\mathrm{ob}}:\;\;\Sigma _{\mathrm{ob}}\left( x,y\right) =0
\label{h4.1}
\end{equation}
The boundary conditions are formulated as follows. 
\begin{equation}
\begin{array}{c}
\mathcal{S}_{1}:\;\psi \left( x,-R\right) =\psi _{1}=\text{const,\qquad }%
\mathcal{S}_{2}:\;\psi \left( x,R\right) =\psi _{2}=\text{const} \\ 
\mathcal{S}_{\mathrm{ob}}:\;\psi \left( x,R\right) =\psi _{\mathrm{ob}}=%
\text{const}
\end{array}
\label{h4.2}
\end{equation}
Constants $\psi _{1}$ and $\psi _{2}$ are connected by the relation 
\begin{equation}
\psi _{2}-\psi _{1}=-\int\limits_{-R}^{R}V_{x}\left( y\right) dy
\label{h4.3}
\end{equation}
Conditions (\ref{h4.2}) mean that the normal velocity component vanishes on
all walls. Value of $\psi _{\mathrm{ob}}$ describes a circulation around the
obstacle, because $\psi _{\mathrm{ob}}-\psi _{1}$ describes the mean value
of flux between $\mathcal{S}_{1}$ and $\mathcal{S}_{\mathrm{ob}}$. For
instance, if $\psi _{\mathrm{ob}}=\psi _{1}$, the circulation around the
obstacle is such, that the fluid flows only between $\mathcal{S}_{2}$ and $%
\mathcal{S}_{\mathrm{ob}}$

Boundary condition on the inflow surface $\mathcal{S}_{\mathrm{in}}$ has the
form 
\begin{equation}
\mathcal{S}_{\mathrm{in}}:\;\psi \left( -L,y\right) =\psi
_{1}-\int\limits_{-R}^{y}V_{x}\left( y\right) dy  \label{h4.4}
\end{equation}
where the velocity component $V_{x}\left( y\right) $ along the $x$-axis is
presumed to be given. The boundary condition on the outflow surface $%
\mathcal{S}_{\mathrm{out}}$ has the form 
\begin{equation}
\mathcal{S}_{\mathrm{out}}:\;\psi \left( L,y\right) =\psi _{\mathrm{out}%
}\left( y\right) ,\qquad \psi _{\mathrm{out}}\left( -R\right) =\psi
_{1},\qquad \psi _{\mathrm{out}}\left( R\right) =\psi _{2}  \label{h4.5}
\end{equation}
where $\psi _{\mathrm{out}}\left( y\right) $ is some given function. It
describes the velocity distribution on the outflow surface. To understand,
why such a distribution should be given, and why it is not determined by the
inflow condition, let us imagine that outside the volume $V$ in vicinity of $%
\mathcal{S}_{\mathrm{out}}$ there are some obstacles. Distribution of
velocity on $\mathcal{S}_{\mathrm{out}}$ depends on size, shape and position
of these obstacles. Description of these obstacles is carried out by $\psi _{%
\mathrm{out}}\left( y\right) $.

Let us consider a flow over the obstacle and rewrite dynamic equation (\ref
{h2.32}) in the form 
\begin{equation}
\mathcal{L}\psi =\Omega \left( \psi \right)  \label{h4.6}
\end{equation}
\begin{equation}
\mathcal{L}\psi =\nabla ^{2}\psi =\psi _{,11}+\psi _{,22}  \label{h4.8}
\end{equation}
\begin{equation}
\Omega \left( \psi \right) =-\rho _{0}\Omega ^{12}(\xi _{2})=-U\left( \xi
_{2}\right) =-U\left( Q\left( \psi \right) \right)  \label{h4.9}
\end{equation}
where the function $U$ is defined as a function of argument $y$ by means of
the relation (\ref{h3.5}) 
\[
U\left( y\right) =\frac{\partial }{\partial y}V_{x}\left( y\right) -W\left(
y\right) 
\]
Function $\xi _{2}=Q\left( \psi \right) $ is defined as a function of
argument $\psi $ by means of the relation (\ref{h4.4}) 
\begin{equation}
\psi =\psi _{1}-\int\limits_{-R}^{Q\left( \psi \right) }V_{x}\left( y\right)
dy  \label{h4.10b}
\end{equation}
In particular, if $V_{x}\left( y\right) =V_{0}=$const, $\xi _{2}=Q\left(
\psi \right) =\left( \psi _{1}-\psi \right) /V_{0}$.

Dynamic equation (\ref{h4.6}) is a nonlinear equation, in general. Its
solution is difficult. The goal of this paper is to provide a statement of
the stream flow problem, because even this problem is not solved for
irrotational flows. So far as the solution of equation (\ref{h4.6}) is
concerned, it is a special mathematical problem which can be solved by
different methods.

If vorticity $\Omega $ is small, dynamic equation (\ref{h4.6}) can be solved
by means of an iteration procedure. To obtain the zeroth order approximation 
$\psi _{\left( 0\right) }$, we consider linear equation 
\begin{equation}
\mathcal{L}\psi _{\left( 0\right) }=\nabla ^{2}\psi _{\left( 0\right) }=0
\label{h4.10}
\end{equation}
with boundary conditions (\ref{h4.2}) -- (\ref{h4.5}), which is written for
brevity in the form 
\begin{equation}
\left[ \psi _{\left( 0\right) }\right] _{\Sigma }=\psi _{\Sigma }\left( t,%
\mathbf{x}\right) ,\qquad \mathbf{x}\in \Sigma  \label{h4.11}
\end{equation}
where $\Sigma $ is the boundary of the volume filled by fluid.

Let $G\left( \mathbf{x|x}^{\prime }\right) $ be the Green function for
equation (\ref{h4.10}). It is defined by the relations 
\begin{equation}
\nabla ^{2}G\left( \mathbf{x|x}^{\prime }\right) =-4\pi \delta \left( 
\mathbf{x-x}^{\prime }\right) ,\qquad G\left( \mathbf{x|x}^{\prime }\right)
_{x\in \Sigma }=0  \label{h4.12}
\end{equation}
Solution of equation (\ref{h4.10}) is given by the formula 
\begin{equation}
\psi _{\left( 0\right) }\left( \mathbf{x}\right) =-\frac{1}{4\pi }%
\int\limits_{\Sigma }\frac{\partial G\left( \mathbf{x|x}^{\prime }\right) }{%
\partial n^{\prime }}\psi _{\Sigma }\left( x^{\prime }\right) dS^{\prime }
\label{h4.14}
\end{equation}
where $\partial /\partial n^{\prime }$ means derivative with respect to $%
x^{\prime }$ in the direction normal to $\Sigma $. Solution (\ref{h4.14}) is
an irrotational approximation to the solution of (\ref{h4.6}) for the
rotational flow.

The first approximation is determined by the relation 
\begin{equation}
\psi _{\left( 1\right) }\left( \mathbf{x}\right) =\psi _{\left( 0\right)
}\left( \mathbf{x}\right) -\frac{1}{4\pi }\int\limits_{V}G\left( \mathbf{x}|%
\mathbf{x}^{\prime }\right) \Omega \left( \psi _{\left( 0\right) }\left( 
\mathbf{x}^{\prime }\right) \right) dV^{\prime }  \label{h4.15}
\end{equation}
where integration is produced over the volume $V$ filled with the fluid. 
\begin{equation}
\psi _{\left( n\right) }\left( \mathbf{x}\right) =\psi _{\left( 0\right)
}\left( \mathbf{x}\right) -\frac{1}{4\pi }\int\limits_{V}G\left( \mathbf{x}|%
\mathbf{x}^{\prime }\right) \Omega \left( \psi _{\left( n-1\right) }\left( 
\mathbf{x}^{\prime }\right) \right) dV^{\prime },\qquad n=2,3,...
\label{h4.16}
\end{equation}
If the iteration process converges, we obtain a solution of the problem of
rotational stationary flow around an obstacle. One can hope that the
iteration process appears to be effective at least for small vorticity $%
\Omega $.

\section{Examples of two-dimensional stationary flow.}

Let us consider a stationary flow, bounded by two parallel plane walls $%
\mathcal{S}_{1}:\;y=0,\qquad \mathcal{S}_{2}:\;y=b$, with inflow and outflow
surfaces $\mathcal{S}_{\mathrm{in}}:\;x=0,\qquad \mathcal{S}_{\mathrm{out}%
}:\;x=a$. To obtain linear equation (\ref{h4.6}) which has a simple exact
solution, we choose the following boundary condition 
\begin{eqnarray}
\psi \left( x,0\right) &=&0,\qquad \psi \left( x,b\right) =0,  \label{h5.3a}
\\
\psi \left( 0,y\right) &=&V_{0}\sin \frac{\pi y}{b},\qquad \psi \left(
a,y\right) =V_{0}\sin \frac{\pi y}{b},\qquad V_{0}=\text{const}
\label{h5.3b} \\
\psi _{,11}\left( 0,y\right) &=&k^{2}V_{0}\sin \frac{\pi y}{b},\qquad k=%
\sqrt{\Omega +\left( \frac{\pi }{b}\right) ^{2}},\qquad \Omega =\text{const}
\label{h5.3c}
\end{eqnarray}
In this case the vorticity $\Omega \left( \psi \right) $ in the rhs of (\ref
{h4.6}) on the boundary $x=0$ is determined by the first condition (\ref
{h5.3b}) and (\ref{h5.3c}) 
\begin{eqnarray*}
\frac{\partial }{\partial y}u\left( 0,y\right) -\frac{\partial }{\partial x}%
v\left( 0,y\right) &=&\psi _{,11}\left( 0,y\right) +\psi _{,22}\left(
0,y\right) =\left( k^{2}-\left( \frac{\pi }{b}\right) ^{2}\right) V_{0}\sin 
\frac{\pi y}{b} \\
&=&\Omega \psi \left( 0,y\right)
\end{eqnarray*}
It means that the function $\Omega \left( \psi \right) $ in the rhs of (\ref
{h4.6}) has the form $\Omega \left( \psi \right) =\Omega \psi $,$\;\;\Omega
= $const, and dynamic equation takes the form of linear equation 
\begin{equation}
\nabla ^{2}\psi =\Omega \psi ,\qquad \Omega =\text{const}  \label{h5.2}
\end{equation}
Its solution, satisfying the boundary conditions (\ref{h5.3a}), (\ref{h5.3b}%
) has the form 
\begin{equation}
\psi =V_{0}\frac{\sinh \left( kx\right) +\sinh \left( k\left( a-x\right)
\right) }{\sinh \left( ka\right) }\sin \frac{\pi y}{b},\qquad k=\sqrt{\Omega
+\left( \frac{\pi }{b}\right) ^{2}}  \label{h5.2a}
\end{equation}
This solution describes a set of flows with vanishing total flux through the
inflow surface and different vorticities $\Omega $. Flows of this set are
distinguished by the power of circulation which is described by the
parameter $\Omega $.

Let us consider two-dimensional stationary slightly rotational flow between
two concentric circular cylinder of radius $a$ and $b$ $(a<b)$. Let us
choose the coordinate system with the $z$-axis directed along the cylinder
axis and coordinate origin on the cylinder axis. The $x$-axis is directed
along the mean velocity $V_{0}$ of the flow. Let us introduce polar
coordinates $\left( r,\varphi \right) $ in the $\left( x,y\right) $-plane
and set the boundary conditions for the stream function as follows. 
\begin{eqnarray}
\psi \left( a,\varphi \right) &=&0,\qquad \psi \left( b,\varphi \right)
=-V_{0}\left( 1-\frac{a^{2}}{b^{2}}\right) b\sin \varphi ,  \label{h6.1} \\
\left[ \frac{1}{r}\frac{\partial }{\partial r}\left( r\frac{\partial \psi
\left( r,\varphi \right) }{\partial r}\right) \right] _{r=b} &=&\omega
_{1}\left( \varphi \right) =\omega \left( \psi \left( b,\varphi \right)
\right) =\left\{ 
\begin{array}{ccc}
\omega \left( \psi \right) & \text{if} & \left| \psi \right| <\psi _{0} \\ 
0 & \text{if} & \left| \psi \right| >\psi _{0}
\end{array}
\right.  \label{h6.2}
\end{eqnarray}
where $\omega _{1}\left( \varphi \right) $ is some given function of $%
\varphi $. It is supposed that the flow is rotational only in the part of
fluid which flows near the cylinder. Let $-\psi _{0}$ and $\psi _{0}$ 
\begin{equation}
\psi _{0}=V_{0}\left( 1-\frac{a^{2}}{b^{2}}\right) Na  \label{h6.2a}
\end{equation}
be values of the stream function on the boundary between the rotational and
irrotational parts of the flow. $Na$ is the transverse size of the
rotational region $\left( N\gg 1\right) $. After statement of the problem we
set $b\rightarrow \infty $. The irrotational flow of the zeroth
approximation around the cylinder is described by the stream function $\psi
_{\left( 0\right) }$ which is given by the expression \cite{L32} 
\begin{equation}
\psi _{\left( 0\right) }\left( r,\varphi \right) =-V_{0}\left( 1-\frac{a^{2}%
}{r^{2}}\right) r\sin \varphi ,\qquad r>a  \label{h6.3}
\end{equation}
where $r,\varphi $ are polar coordinates on the $\left( x,y\right) $-plane 
\begin{equation}
x=r\cos \varphi ,\qquad y=r\sin \varphi  \label{h6.4}
\end{equation}
The irrotational flow (\ref{h6.3}) is used at formulation of boundary
conditions on the external cylinder of radius $b$.

According to (\ref{h4.15}) and (\ref{h6.1}) - (\ref{h6.3}) we obtain for the
first approximation 
\begin{equation}
\psi _{\left( 1\right) }\left( r,\varphi \right) =\psi _{\left( 0\right)
}\left( r,\varphi \right) -\frac{1}{4\pi }\int \int\limits_{\left| \psi
_{\left( 0\right) }\left( r^{\prime },\varphi ^{\prime }\right) \right|
<\psi _{0}}G\left( r,\varphi |r^{\prime },\varphi ^{\prime }\right) \omega
\left( \psi _{\left( 0\right) }\left( r^{\prime },\varphi ^{\prime }\right)
\right) r^{\prime }dr^{\prime }d\varphi ^{\prime }  \label{h6.5}
\end{equation}
where $\omega \left( \psi \right) $ is defined by the relation (\ref{h6.2}),
and the Green function $G\left( r,\varphi |r^{\prime },\varphi ^{\prime
}\right) $ is taken for the case $b=\infty .$ It has the form 
\begin{equation}
G\left( r,\varphi |r^{\prime },\varphi ^{\prime }\right) =\ln \frac{a^{2}+%
\frac{r^{2}r^{\prime 2}}{a^{2}}-2rr^{\prime }\cos \left( \varphi -\varphi
^{\prime }\right) }{r^{2}+r^{\prime 2}-2rr^{\prime }\cos \left( \varphi
-\varphi ^{\prime }\right) }  \label{h6.6}
\end{equation}
and integration is produced over the region, where $\left| \psi _{\left(
0\right) }\left( r,\varphi \right) \right| <\psi _{0}=$const. According to (%
\ref{h6.2a}), (\ref{h6.3}) this region is defined by the relation 
\begin{equation}
\left| V_{0}\left( 1-\frac{a^{2}}{r^{2}}\right) r\sin \varphi \right|
<V_{0}Na,\qquad r>a,\;\;b=\infty  \label{h6.7}
\end{equation}
where $N\gg 1$. If 
\begin{equation}
r<R=a\left( \frac{N}{2}+\sqrt{\left( \frac{N}{2}\right) ^{2}+1}\right)
\label{h6.8}
\end{equation}
the inequality (\ref{h6.7}) is satisfied for any angle $\varphi $, and for $%
r^{\prime }<R$ one should integrate in (\ref{h6.5}) over all values of angle 
$\varphi ^{\prime }$. For $r^{\prime }>R$ it is useful to introduce the
variable $\eta $ by means of relation 
\begin{equation}
\varphi ^{\prime }=\Phi \left( r,\eta \right) =\arcsin \frac{\eta }{%
V_{0}\left( 1-\frac{a^{2}}{r^{2}}\right) r}  \label{h6.9}
\end{equation}
and integrate over $\eta $ in the region, where $r^{\prime }>R$.

We obtain 
\begin{equation}
\psi _{\left( 1\right) }=\psi _{\left( 0\right) }+I_{1}+I_{2}+I_{3}
\label{h6.10}
\end{equation}
where 
\begin{equation}
I_{1}=-\frac{1}{4\pi }\int\limits_{a}^{R}r^{\prime }dr^{\prime
}\int\limits_{0}^{2\pi }G\left( r,\varphi |r^{\prime },\varphi ^{\prime
}\right) \omega \left( \psi _{\left( 0\right) }\left( r^{\prime },\varphi
^{\prime }\right) \right) d\varphi ^{\prime }  \label{h6.11}
\end{equation}
\begin{equation}
I_{2}=\frac{1}{4\pi }\int\limits_{R}^{\infty }r^{\prime 2}dr^{\prime
}\int\limits_{-\Phi \left( r^{\prime },\psi _{0}\right) }^{\Phi \left(
r^{\prime },\psi _{0}\right) }\frac{G\left( r,\varphi |r^{\prime },\Phi
\left( r^{\prime },\eta \right) \right) \omega \left( -\eta \right)
V_{0}\left( 1-\frac{a^{2}}{r^{\prime 2}}\right) d\eta }{\sqrt{%
V_{0}^{2}\left( 1-\frac{a^{2}}{r^{\prime 2}}\right) ^{2}r^{\prime 2}-\eta
^{2}}}  \label{h6.12}
\end{equation}
\begin{equation}
I_{3}=\frac{1}{4\pi }\int\limits_{R}^{\infty }r^{\prime 2}dr^{\prime
}\int\limits_{\pi -\Phi \left( r^{\prime },\psi _{0}\right) }^{\pi +\Phi
\left( r^{\prime },\psi _{0}\right) }\frac{G\left( r,\varphi |r^{\prime
},\Phi \left( r^{\prime },\eta \right) \right) \omega \left( -\eta \right)
V_{0}\left( 1-\frac{a^{2}}{r^{\prime 2}}\right) d\eta }{\sqrt{%
V_{0}^{2}\left( 1-\frac{a^{2}}{r^{\prime 2}}\right) ^{2}r^{\prime 2}-\eta
^{2}}}  \label{h6.14}
\end{equation}
where $\psi _{0}=V_{0}Na$. Although calculations of integrals (\ref{h6.11})
-- (\ref{h6.14}) is not simple, they give the first approximation of the
rotational flow around the cylinder.

\section{Conclusion}

Theory of rotational flows is more complicated, than that of irrotational
flows. There are two reasons for this.

\begin{enumerate}
\item  Dynamic equations for rotational flows appear to be nonlinear almost
always. They cannot be solved exactly.

\item  The problem of rotational stream flow is not stated yet properly.
\end{enumerate}

The first problem is pure mathematical. Existence of this problem is quite
clear for all researchers. The second problem is physical. We believe that
hardly anybody of researchers guesses on existence of this problem. The
first problem is very difficult. The second problem is not so difficult, but
it is much more important that the first problem. Why?

First, because existence of the second problem in itself seems to be
problematic. Second, the irrotational flow is a special degenerate case of
the rotational flow, when vorticity vanishes. Transition from the general
case to the degenerate one is rather simple. The reverse transition from the
degenerate case to the general one is much more difficult, because the
general case contains additional information which is absent in the
degenerate case. If we could not evaluate correctly this additional
information, we do not enable to construct a perfect theory. Practically, we
ignore dynamic equations (\ref{b1.4}) and boundary conditions for
determination of vorticity. As a result we obtain imperfect theory, where
some degrees of freedom are ignored. In some cases these hidden degrees of
freedom manifest themselves suddenly, and we try to understand what could be
meant by this effect. At any rate a serious mathematical investigation of
such difficult hydrodynamical problems as turbulence is impossible, if the
underlying rotational flow theory is incomplete.

\end{document}